\documentclass[preprint]{aastex}
\usepackage{graphicx}
\begin{document}

\title{An Enhanced Spectroscopic Census of the Orion Nebula Cluster}
\author{Lynne A. Hillenbrand\altaffilmark{1}}
\email{lah@astro.caltech.edu}
\author{Aaron S. Hoffer\altaffilmark{1,2}}
\email{hofferaa@msu.edu}
\and
\author{Gregory J. Herczeg\altaffilmark{1,3}}
\email{gherczeg1@gmail.com}

\altaffiltext{1}{Department of Astronomy, California Institute of Technology, Pasadena, CA 91125}
\altaffiltext{2}{Current affiliation: Department of Physics and Astronomy, Michigan State University, East Lansing, 48823}
\altaffiltext{3}{Current affiliation: Kavli Institute for Astronomy and Astrophysics, Peking University; Yi He Yuan Lu 5, Hai Dian Qu; Beijing 100871, P. R. China}

\begin{abstract}
We report new spectral types or spectral classification constraints for over 600 stars in the Orion Nebula Cluster (ONC) based on medium resolution ($R\approx 1500-2000$) red optical spectra acquired using the Palomar 200$"$ and Kitt Peak 3.5m telescopes.  Spectral types were initially estimated for F, G, and early K stars from atomic line indices while for late K and M stars, constituting the majority of our sample, indices involving TiO and VO bands were used.  To ensure proper classification, particularly for reddened, veiled, or nebula-contaminated stars, all spectra were then visually examined for type verification or refinement. We provide an updated spectral type table that supersedes \citet{1997AJ....113.1733H}, increasing the percentage of optically visible ONC stars with spectral type information from 68\% to 90\%.  However, for many objects, repeated observations have failed to yield spectral types primarily due to the challenges of adequate sky subtraction against a bright and spatially variable nebular background.  The scatter between our new and our previously determined spectral types is approximately 2 spectral sub-classes.  We also compare our grating spectroscopy results with classification based on narrow-band TiO filter photometry from \citet{DaRio10,DaRio12}, finding similar scatter.  While the challenges of working in the ONC may explain much of the spread, we highlight several stars showing significant and unexplained bona fide spectral variations in observations taken several years apart; these and similar cases could be due to a combination of accretion and extinction changes.  Finally, nearly 20\% of ONC stars exhibit obvious \ion{Ca}{2} triplet emission indicative of strong accretion. 
\end{abstract}

\section{Introduction}
\subsection{The Orion Nebula Cluster}
Located at less than one-half kpc from the Sun and 
at moderate galactic latitude ($b\approx -20$), the greater Orion region 
is the nearest example of recent and ongoing massive star formation; 
see \citet{2008hsf1.book..459B} for an overview. 
A subclustered OB association extends over many tens of pc 
and contains stars up to a few tens of Myr old. To the southeast, the  
Orion A and B molecular cloud complex harbors stellar nurseries each
only a fraction of a pc to a few pc in size, and less than one Myr old.
The Orion composite is a revered touch stone for our developing understanding 
of star and planet formation processes on this range of 
spatial and temporal scales.  The identification and detailed study
of the young stellar population across the Orion complex is our best opportunity
for probing in three dimensions a resolved star formation history.  

Part of the so-called ``integral shaped filament" of molecular gas 
\citep{1987ApJ...312L..45B}, the northern part of the Orion A cloud 
has up to 100 magnitudes of total visual extinction through its densest regions. 
The Orion Nebula, also known as M42 or NGC 1976, betrays a rich young stellar cluster, 
the Orion Nebula Cluster (ONC), the center of which contains  
the massive Trapezium stars \footnote{Traditionally, the Trapezium Cluster is
the densest part of the ONC within about 2\arcmin\ (corresponding to $\sim$0.3 pc), while the larger
ONC extends to about 20\arcmin\ ($<$3pc)}. 
The current three-dimensional model of the Orion Nebula indicates 
a thin shell of ionized gas \citep{1995ApJ...438..784W} with the main ionizing source 
region $\theta^{1}$ Ori C.  Another $\sim$20 stars A0 or earlier 
are within the HII region as is a sizable population (many thousand members)
of later spectral type T Tauri stars.  These lower mass stars are still 
pre-main sequence and can be used to trace the star formation history as well as measure
the initial mass function (IMF), which in the case of young, unevolved
and co-located stars, is equivalent to the present day mass function. 

In contrast to the sparsely distributed and older OB association, 
the ONC is relatively compact and more recently formed.  Its projection 
on molecular cloud material significantly reduces contamination by background stars.
However, the high and spatially variable extinction 
\citep[e.g.][]{Scandariato11} renders fewer than 50\% of the $\sim$3500 stars 
within this region having $K < 14$ mag also optically bright, $I < 17.5$ mag.  
Nevertheless, the bulk of the optically visible ONC members are extincted 
by 2 magnitudes or less, with derived extinction values for fainter stars 
ranging up to 8 magnitudes. The deepest optical surveys 
as well as most near-infrared surveys measure more heavily extincted members, 
but are more likely to suffer contamination 
at the faint end -- both from intrinsically bright but highly reddened
background giants seen through the cloud, and from intrinsically faint 
foreground and unreddened but late type field dwarfs. 

Cluster membership can be determined kinematically in the ONC.  Using astrometric
techniques, \citet{1988AJ.....95.1755J} identified 891 probable members 
from 997 measured proper motions (measured among 1052 optically visible stars). 
Radial velocities \citep[e.g.][]{Tobin09} have also been used to establish 
cluster membership.  High precision three-dimensional space motion data from these
and other ongoing investigations can be
combined with precise cluster distance estimates 
\citep{2007MNRAS.376.1109J,
2007A&A...466..649K,
2007PASJ...59..897H,
2007ApJ...667.1161S,
2007A&A...474..515M},
to study the cluster structure and dynamics and test theories 
for formation and evolution of dense clusters in molecular clouds. 
Additional observation and analysis work in this area is needed.

The ONC contains a broadly populated stellar mass function extending from
a single massive (late-O type) star all the way down to low-mass stars
(having late M types) and into the sub-stellar mass regime of brown dwarfs 
(likely L and T type), as well as planetary mass objects remaining to be spectrally typed
and identified as such.
With a mean stellar age for the optically visible stars of less than 2 Myr \citep{1986ApJ...307..609H,1997AJ....113.1733H,DaRio10,Reggiani11}, the ONC is not only one of the largest but also one of the densest ($\sim 10^4$ stars pc$^{-3}$ in the inner cluster; \cite{1994AJ....108.1382M,1998ApJ...492..540H}) nearby star forming regions.  The ONC population was used to demonstrate definitively 
that it is possible to form both high and low mass stars as well as 
brown dwarfs in the same $<$0.1 pc region. 
From the lower mass stars, the estimated 
current star formation rate is $\sim 10^{-4} M_{\sun}$ per year
\citep[][hereafter H97]{1997AJ....113.1733H}.

The H97 study of the ONC employed optical 
spectroscopy and photometry to individually de-redden and locate over 900 
stars on the H-R diagram, enabling estimation of their masses and ages 
via comparison to pre-main sequence evolutionary tracks. Work by 
\citet{2000ApJ...540.1016L},
\citet{2004ApJ...610.1045S}, \citet{2007MNRAS.381.1077R} and
\citet{2009MNRAS.392..817W} increased 
the spectroscopically studied sample, employing optical and near-infrared spectral
classification techniques.  
\citet{DaRio10} and \citet{DaRio12} estimated spectral types for M-type stars 
from narrow band photometry, as discussed in more detail below. 
Additional studies such as those of
\citet{HC00}, \citet{2002ApJ...573..366M}, \citet{2005MNRAS.361..211L}, and
\citet{2011A&A...534A..10A} used photometric techniques rather than spectroscopic, 
and arrived at IMF estimates that are useful in a statistical
sense but do not determine masses and ages for individual stars.
The general consensus from the above studies is that the ONC IMF rises from
the highest masses to sub-solar masses, realizes a flattening in the
$0.5-0.6 M_{\sun}$ regime, and peaks in the $0.2-0.3 M_{\sun}$ regime before
turning over throughout the sub-stellar (brown dwarf) mass range. 
There is some discrepancy as to the exact form of the IMF below the peak and
turnover in this cluster (and in many other young clusters).
Increased spectroscopic samples as well as improved understanding 
of contraction models at young ages will be required 
for progress on the IMF in the low mass stellar and sub-stellar domain.

Recent reviews of the ONC region are those by \citet{2008hsf1.book..483M} on the stellar population and historical distance measurements, and by \citet{2008hsf1.book..544O} on the local ONC interstellar medium including circumstellar structures.  
The ONC stellar population has been the sample of choice for investigations 
of various stellar and circumstellar phenomena, in addition to the IMF
studies described above, in part because it is the nearest and youngest 
example of an entire stellar/sub-stellar mass spectrum.
Recent studies of circumstellar dust and gas
include those of \citet{Megeath12},
\citet{Mann09} and \citet{2008AJ....136.2136R}, 
of disk accretion include 
\citet{2012ApJ...755..154M} and \citet{2005AJ....129..363S},
and of variability-disk connections include \cite{2012ApJ...753..149M}.
Recent studies on stellar properties include topics such as
rotation and disk-rotation connections \citep{
2001AJ....122.3258R, 2001AJ....121.3160C,
2002A&A...396..513H, 2005ApJ...633..967H, 2006ApJ...646..297R,
2007ApJ...671..605C}, 
lithium depletion \citep{2007ApJ...659L..41P},
coronal activity/flaring 
\citep{2002ApJ...574..258F, 2008ApJ...688..418G, 
2008ApJ...677..401P}, 
stellar abundances \citep{2005ApJ...626..425C},
magnetic fields \citep{2008MNRAS.387L..23P},
and multiplicity \citep[e.g.][]{1999AJ....117.1375S,2007AJ....134.2272R,2006A&A...458..461K}.

\subsection{Motivation for Further Spectroscopic Study}

There is still great interest in improving our knowledge 
of stellar and circumstellar properties within the ONC.  
As described above, the H97 paper and its accompanying database 
have been used in subsequent years by many other investigators studying 
e.g. the stellar/sub-stellar IMF, the age distribution and star formation history, 
stellar/sub-stellar angular momentum evolution,
circumstellar disks and disk evolution, accretion, magnetic activity, lithium,
as well as other properties of young stars in the ONC.  
Recent examples of major surveys of the ONC  with forefront facilities that 
reach unprecedented depth across the electromagnetic spectrum, 
include those from X-ray (with Chandra; PI E. Feigelson) 
to optical (with Hubble; PI M. Robberto)
to mid-infrared (with Spitzer; \citet{Megeath12}) and finally
to millimeter (with the SMA and CARMA interferometers).
Synoptic photometric monitoring studies by various groups at optical and infrared wavelengths  
\citep[e.g.][]{2011ApJ...733...50M} continue.  

Nearly all of these previous analyses and new surveys are limited in their analysis phases,
however, due to lack of completeness of the ONC spectroscopic survey.
Critically important is that due to 
the wide range in line-of-sight extinction values to individual stars,
spectra are necessary for de-reddening and thus derivation of
extinction estimates, and for subsequent placement on the HR diagram
from which stellar quantities such as stellar luminosity, stellar radius, 
and (further guided by theory) stellar age and mass, angular momentum, etc.

The optical color-magnitude diagram for the ONC suggests that the H97 
optical photometric database is sensitive down to the hydrogen burning limit 
for stars less obscured than 2 magnitudes of visual extinction and
younger than 1 Myr \citep[e.g.][H97]{DaRio10}.
Infrared surveys extend well into the brown dwarf regime with
the stellar-substellar boundary identified at spectral type M6.5 
given the age of the ONC.  
While several thousand young objects are known,
only about 63\% of the $\sim$1500 stars with $I <17$ mag  
had spectral types following the H97 study (and only about 5\% before).
Considering both optical and infrared-only sources in the region,
less than 25\% of the total number of known sources have had 
their spectral types determined.  While infrared spectroscopy 
will be required in order to obtain a complete spectroscopic census of 
the ONC stellar population, optical spectral types can be obtained 
for several hundred more objects than have been published thus far.

The present work aims to enhance the H97 optical spectroscopic census. 
By completing our optical study of the stellar population,
it will be possible for other authors 
to revisit interests in many of the above goals.

\subsection{Optical Spectral Typing and Challenges in the ONC}
Spectral type is a key characteristic of any star, and spectral typing 
is a typical first step in 
estimating the stellar effective temperature.  This can lead, in 
combination with other data or assumptions, to derivation of other
stellar quantities.
The optical wavelength range is important for spectral typing because it 
contains a rich set of molecular and atomic absorption features that are 
temperature dependent.  M type stars, which dominate the mass function,
are characterized by metallic oxide TiO and VO bands. 
These molecular features diminish at hotter temperatures with 
mid-K and earlier stars classified using atomic line strengths and ratios
involving e.g. \ion{Ca}{1} and \ion{Fe}{1}.  
Reliable surface gravity signatures include the shapes of some of the
molecular bands as well as the strength of the Na I doublet 
($\lambda\lambda$ 8183, 8194 \AA) for spectral types later than M2.
For earlier type stars, most reliable surface gravity indicators lie 
blueward of our spectral coverage. 

Relevant to the ONC, spectral types can be determined largely independent 
of reddening and nebular emission when these effects are modest, 
though in extreme cases they do bias or even hinder the analysis.
For late type spectra, the ratios of the depths of 
metal oxide wavebands are not significantly affected by low reddening 
and the possibility of reddening can be taken into account 
when assigning types.
Nebular emission, specfically from low Balmer (e.g. H$\alpha$)
and high Paschen lines as well as the Paschen continuum can also affect 
the spectral typing process or even prevent it in extreme cases, 
especially for faint objects or if background subtraction 
(a combination of terrestrial sky and ONC nebular backgrounds)
is ineffective.  For late type stars the TiO and VO bands 
are generally strong enough such that the effects of minor nebular emission 
can be overcome.  For earlier spectral type stars, however,
strong nebular contamination is more problematic for spectral typing.

A final effect is that of non-nebular continuum and emission lines that are 
associated in young stars with disk-to-star accretion processes.
Spectral typing can be influenced by non-photospheric contributions
to emergent spectrum and in extreme cases the spectrum is accretion-dominated. 

\section{New Optical Spectroscopic Observations}
We obtained low-resolution optical spectra in 15 configurations
with the HYDRA multi-object spectrograph on the 3.5m WIYN telescope
at Kitt Peak during two runs in 2006-2007. We also used the
(now decomissioned) Norris multi-object spectrograph
on the 5m Hale Telescope at Palomar in 1999 to obtain 1 configuration 
on the ONC at the very end of a night allocated to another program.
HYDRA consists of a 1024x2048 CCD fed with 99 fibers, each with a
$2^{\prime\prime}$ aperture (Barden \& Armandroff 1995).
The field size is 60\arcmin\ in diameter.
We used the 316 l/mm grating blazed at 7500 \AA\
setting of the bench spectrograph and the GG-495 filter to
obtain spectra from 5000-10000 \AA\ at $R\sim1500$.
Norris consisted of a 2048x2048 CCD illuminated with 176 fibers, each with a
$1.^{\prime\prime}5$ aperture (Hamilton et al. 1993).  The field size was 20\arcmin\
in diameter.  Our Norris observations were obtained with the 600 l/mm grating
blazed at 5000 \AA\ to produce 54 \AA/mm dispersion (1.3 \AA/pixel).  
The spectral range is 6100--8750 \AA\ at $R\sim2000$. 

Table 1 contains the observation log.
Fiber assignments were prioritized based on ancillary information
such as existence of a published rotation period, exhibition of
large amplitude variability, and proper motion
membership probability; however, our data set is sizable and deep enough 
that these original prioritization criteria are not a significant bias.
Both instruments were configured to maximize the total number of 
quality science targets and minimize the number of repeated spectra. 
However there are many cases of multiple observations of the same object.  
This is useful mitigation against sky subtraction challenges 
and also provides independent assessment of spectral types.

In each fiber configuration, between 30--120 fibers were assigned to a
stellar position and between 13--54 fibers were assigned to sky positions.
The set of observations for each configuration includes a long
exposure on-target, a shorter sky exposure offset
$6^{\prime\prime}-10^{\prime\prime}$ from the target position, and a
set of dome flats and comparison lamps exposures.  The nearby sky exposure
permits accurate correction for nebulosity, which is spatially variable
and hence not accurately represented in averaged spectra
of all in-field sky fibers.  During our Norris
observations we obtained three on-target and one offset sky integration. 
During our HYDRA observations the number of on-target
and sky position integrations varied. Total exposure times are listed in Table 1.

\section{Raw Data Processing and Spectral Extraction}
The observations were reduced using custom routines written in {\it
IDL}.  The bias level was corrected in each image using the CCD overscan
region and the flatfield correction appropriate to each fiber was applied.
The trace of each fiber on the detector was determined
using the dome flat with the absolute position of each fiber 
independently derived for each individual exposure.
Cosmic rays were corrected by identifying deviations from the
illumination profile across each fiber on the detector.  
Scattered light was accounted for by fitting a spline to pixels between
fiber positions.  The counts in each fiber were re-sampled onto
a sub-pixel scale  across the dispersion axis to ensure that the
extraction window remained constant for each separate on-target, sky,
and flat exposure within a given configuration.  
Background counts were subtracted for each fiber based on the counts
between fiber positions.  

The spectrum from each fiber was extracted using a window approximately 1.5 times
the average fiber FWHM on the detector, or 5 pixels 
in the HYDRA spectra and 6 pixels 
in the Norris spectra.  The 2nd-order wavelength solution 
(a FeAr lamp for the Norris observations and a CuAr lamp for the HYDRA observations)
was calculated independently for each fiber. 
We obtained final spectra by summing the counts extracted for
a given fiber from each on-target integration and correcting for sky and
nebular emission by subtracting the counts extracted for the same fiber in the sky
exposure, scaled to the difference in observing times.  In several
configurations the sky emission was scaled by an additional 10--20\% 
to account for changes in the sky transmission.  A master sky spectrum 
was created for each fiber configuration by combining the spectra
obtained from all fibers assigned to the sky, accounting for
fiber-to-fiber sensitivity differences.  Changes in the 
sky emission could be accounted for by comparing 
the master sky spectrum to that obtained in individual
on-target and sky integrations. 

\begin{figure}
\includegraphics[scale = .6,angle=-90]{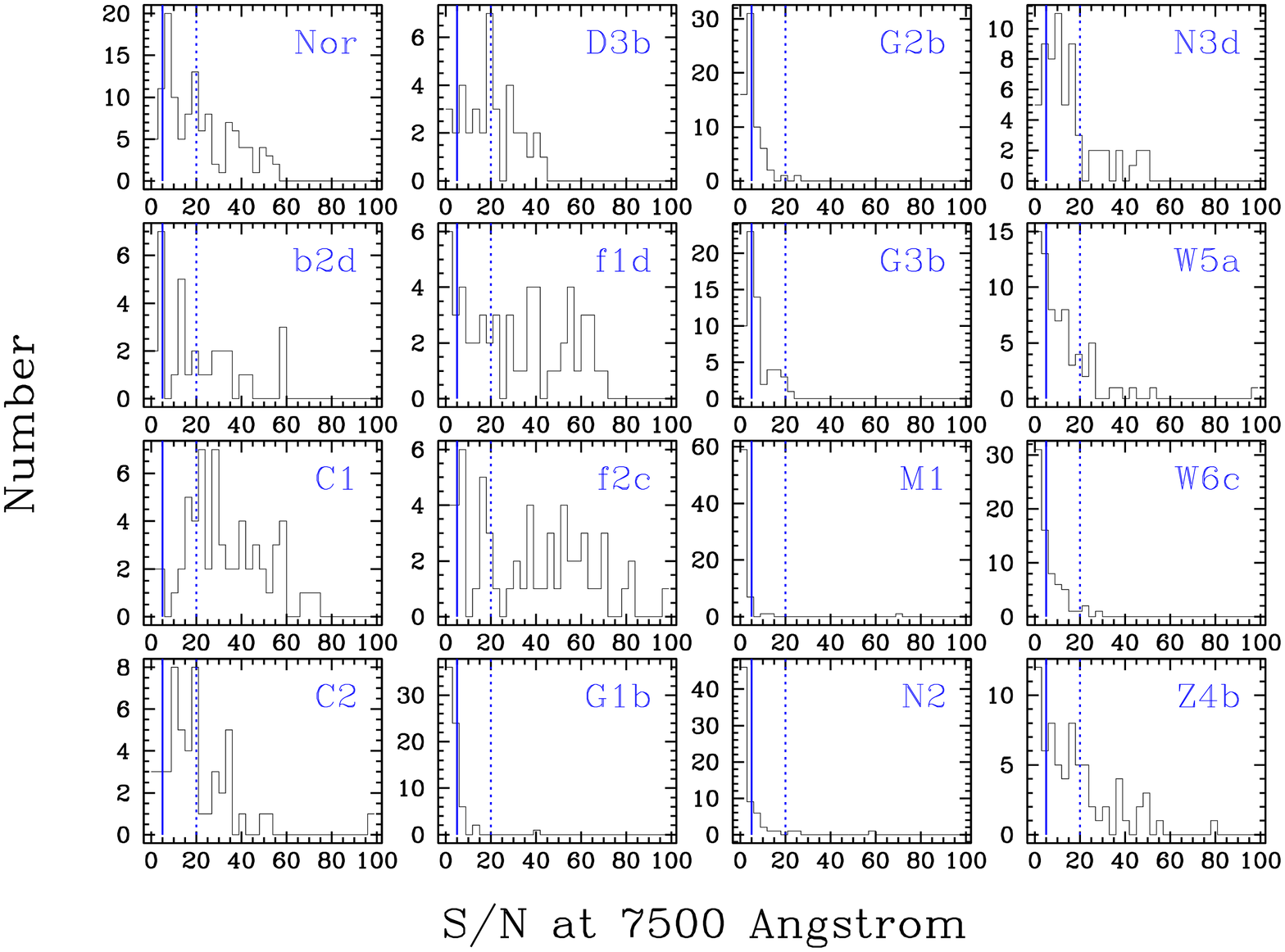}
\caption{Histograms of the achieved signal-to-noise ratio in individual
fiber configurations, with panel labels corresponding to rows in Table 1.
S/N is measured simply as the mean count level 
divided by the root-mean-square of the count level
in the 7500 \AA\ continuum region of each final one-dimensional spectrum.  
Vertical lines drawn at S/N values of 5 (solid) and 20 (dotted) are to guide 
the eye in comparing the panels.  The fiber data sets have varying 
levels of overall quality, mostly due to weather issues.
\label{fig:snr}}
\end{figure}

Figure~\ref{fig:snr} 
shows the distribution of signal-to-noise for the data set,
separated by fiber configuration.  Despite the significant variation in 
data quality among the configurations (due to weather; see Table 1), 
overlays of spectra of the same object repeated in different fiber 
configurations show remarkable agreement in the continuum slope.  
That stated, some configurations are generally poor (e.g. M1) while others 
(such as f2c and f1d) are quite good and these quality factors are
taken into account when assigning final spectral types.

In total we obtained 936 spectra, of which 707 were unique objects 
and 229 were duplicates of the same star obtained in more than one 
fiber configuration.  In some cases the duplication
was intentional due to poor signal-to-noise in the first obtained spectrum.
In other cases, fibers that could not be assigned to previously unobserved 
targets were assigned to stars with existing spectra rather than
random sky positions.
When we have more than one spectrum, we typically have 2,
but in a few cases 3-6 different spectra of the same star were obtained.

\section{Analysis}

Stellar spectral types, spectral type ranges, or spectral type limits
are derivable for 619 unique stars based on photospheric absorption lines.
An additional 88 unique stars were observed, but no spectral type
could be ascertained.
The unclassifiable spectra were either low signal-to-noise ratio 
(the minority) or dominated by contamination 
from bright nebular continuum and/or strong line emission   
(the majority)
that could not be removed using our sky subtraction techniques.
A small fraction, however,
are extreme accreting objects in which the continuum excess 
and line emission is similar to nebular contamination,
but originates in the circumstellar environment.  In cases
of strong \ion{Ca}{2} triplet emission the two possibilities can be distinguished,
but in neither scenario can an underlying spectral type be determined. 
Consistent with convention, the objects lacking absorption features
but having strong \ion{Ca}{2} triplet emission are noted as ``cont$+$emis".

\begin{figure}
\includegraphics[scale = .5]{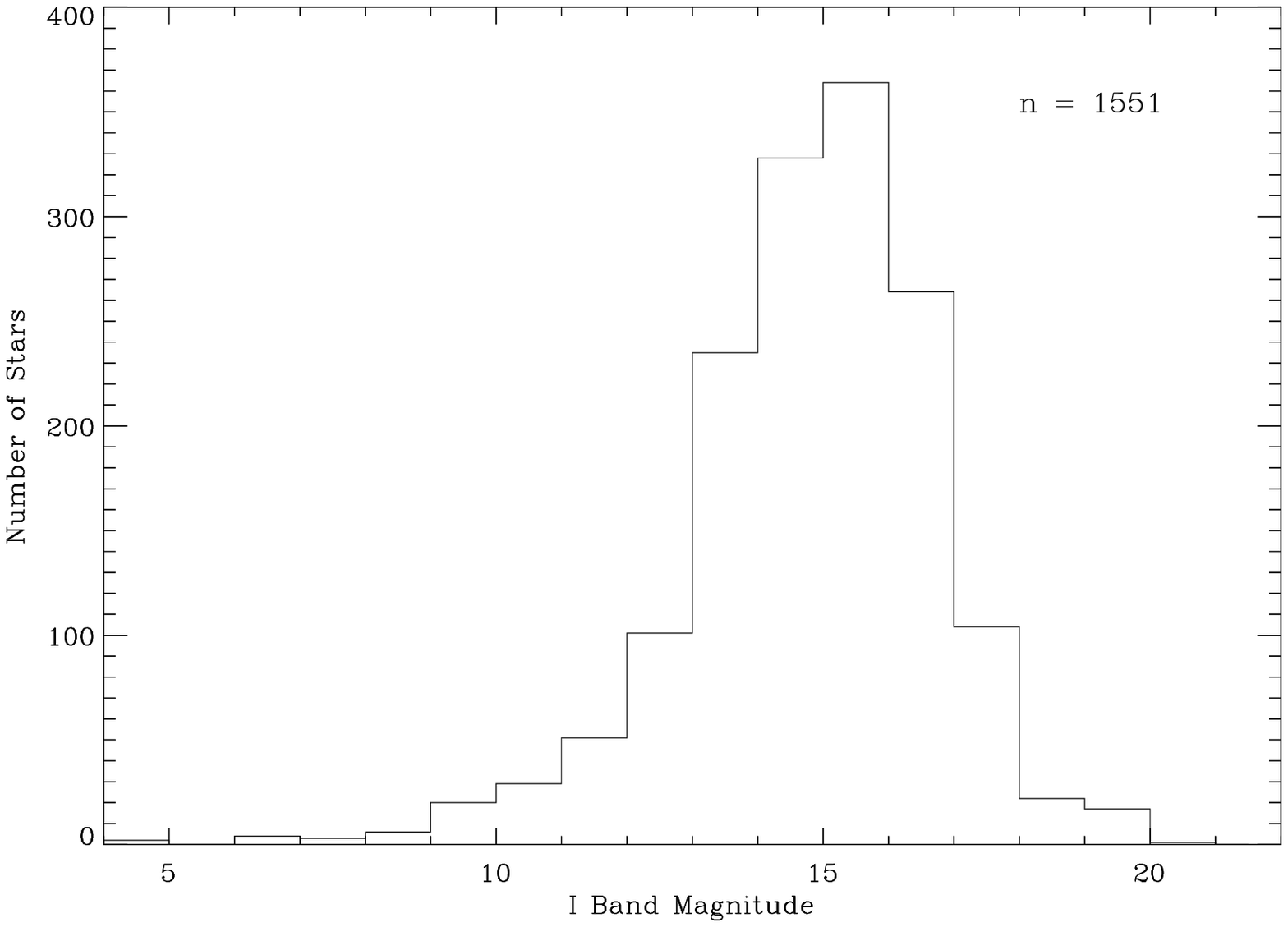}
\\
\includegraphics[scale = .5]{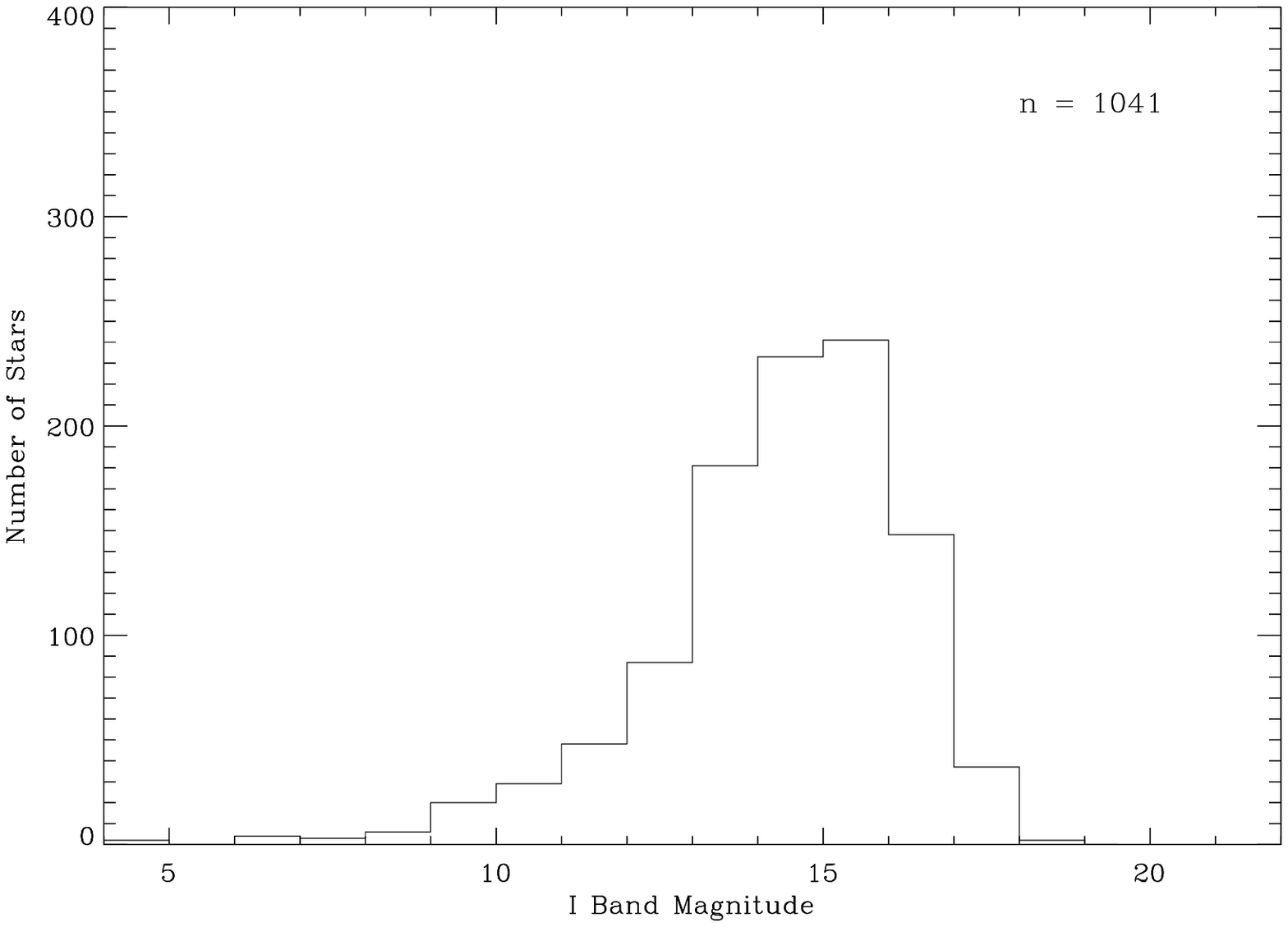}
\\
\includegraphics[scale = .5]{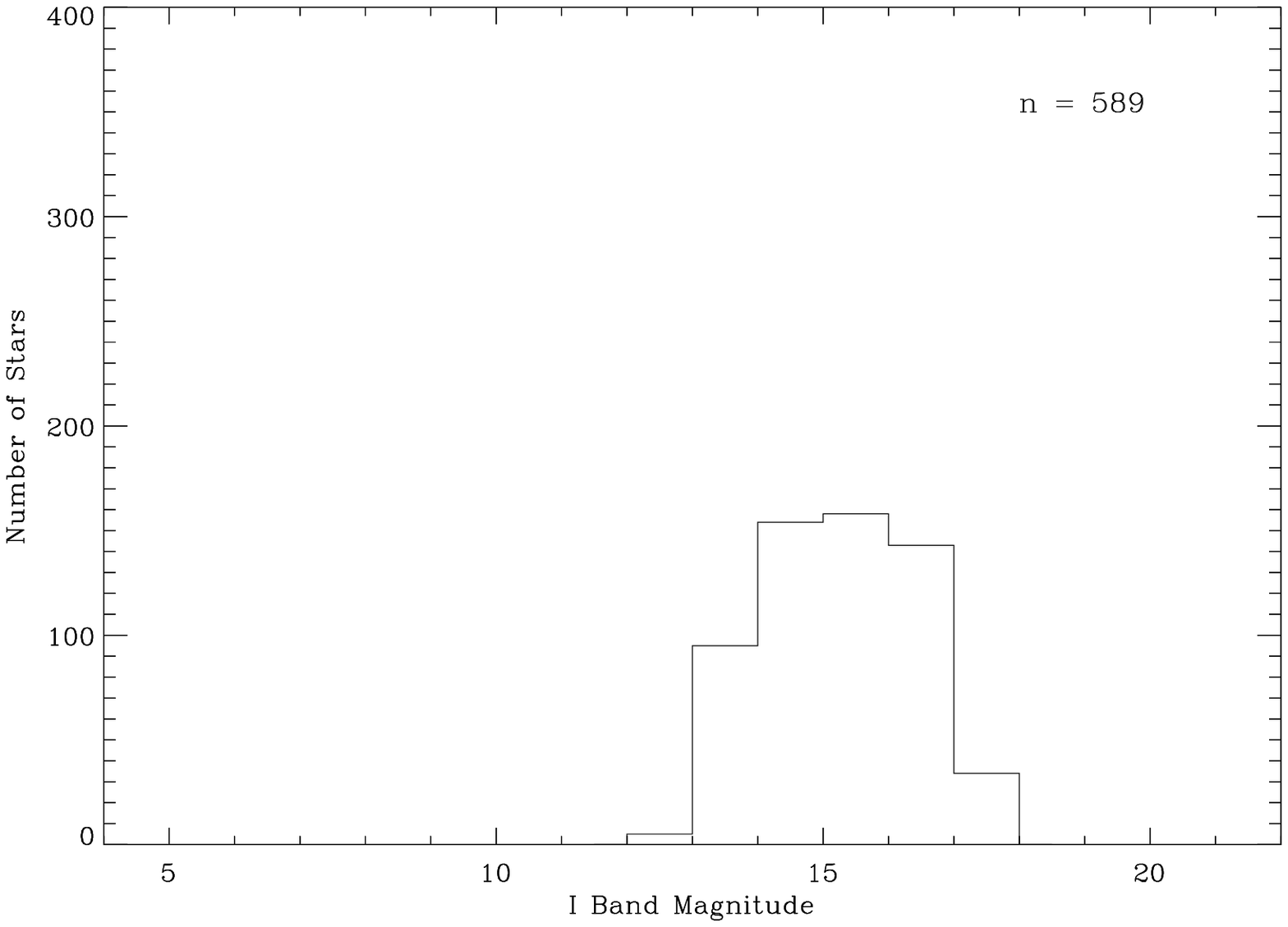}
\caption{
Histograms of I-band magnitudes for the H97 sample,
with the top panel showing the full photometric sample, the middle panel
highlighting those stars with spectral types available at the time of
that publication, and the bottom panel showing those stars with
spectral types (excluding limits) presented in the last column of Table 2. 
There is overlap between middle and bottom panels.
\label{fig:ihist}}
\end{figure}

Figure~\ref{fig:ihist} shows distributions of I-band magnitudes
taken from H97 for all stars analyzed in that paper, as well as for the subset
that had spectral types at that time, and also for those that are spectral typed 
in this work.   The distributions are somewhat similar but
the new set of spectra is biased towards the fainter end of the magnitude distribution. 
In this section we discuss the spectral classification process including
its sensitivity to effective temperature and surface gravity (Section 4.1),
then our quality control procedures (Section 4.2).
Our results are presented in Table 2 (Section 4.3).
Finally we discuss the \ion{Ca}{2} emitters which exhibit unambiguous evidence for
disk-to-star accretion (Section 4.4).

\subsection{Spectral Classification Procedure}

To classify the spectra 
we follow the same procedures as are described in Section 3.3 of H97.
Specifically, the classification was performed 
using spectral indices to measure feature depths,
in the spirit of O'Connell (1973), and followed up by visual examination 
of the data.  For each feature, 30 \AA\ wide bandpasses were used in creating 
spectral strength indices from defined on-feature and off-feature wavelength 
ranges;  average fluxes over these bands were calculated.
\citet{2007MNRAS.381.1067R} give a thorough description of 
the relevant molecular and atomic features and their behavior with
temperature and surface gravity, as well as the challenges
and pitfalls in classifying young stars associated with star formation regions.
These effects include accounting for the observational influence
of low surface gravity,
accretion, large and variable reddening, and nebular contamination in the
classification process.

\begin{figure}
\includegraphics[scale = .9]{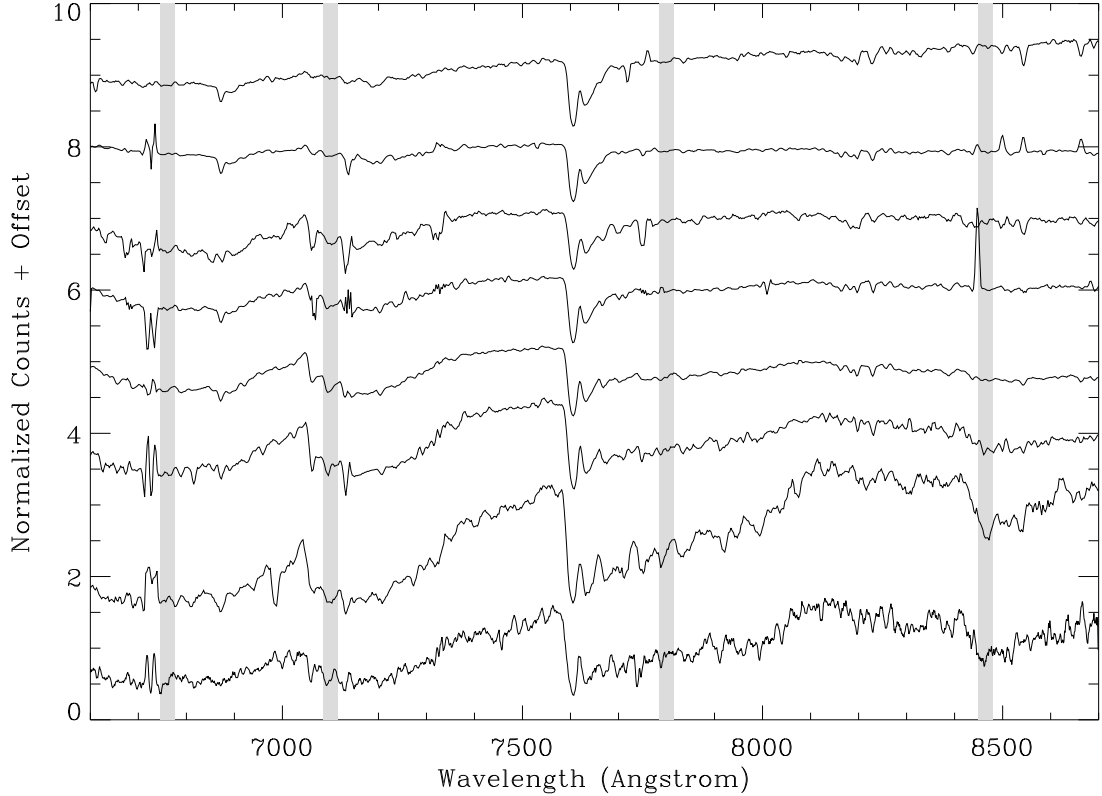}
\caption{Sequence of spectra compared to the 30 \AA-wide regions (shaded) 
used for the TiO band depth measurements in the spectral typing analysis. 
These are among the higher signal-to-noise ratio spectra in our sample, taken
mostly from the WIYN/HYDRA ``f2c" configuration.
The objects and their spectral types are (top to bottom):  
H97 5170 (K5-K6), JW 762 (K8-M0e), JW 340 (M2), JW 901 (M2.5-M3.5), H97 5084 (M4), JW 812 (M5), H97 3123 (M6), and JW 1036 (M8-M9). 
Note the over- and/or under-subtraction in various spectra
of numerous nebular emission lines, notably H$\alpha$.
The second spectrum from the top exhibits weak \ion{Ca}{2} triplet emission 
($\lambda\lambda$ 8498 \AA, 8542 \AA, and 8662 \AA) 
as well as \ion{O}{1} emission ($\lambda$ 8446\AA), that
is intrinsic to the source and not nebular contamination.
The top spectrum has the \ion{Ca}{2} lines in absorption.
Note that continuum shapes can be affected by blueing 
due to accretion processes and/or reddening due to circumstellar 
and/or interstellar extinction.
\label{fig:spectra}}
\end{figure}

The vast majority of the stars in the optically visible
ONC sample are young late K and M type stars.
For their classification metal oxide bands were used: 
four of TiO  ($\lambda$6760, $\lambda$7100, $\lambda$7800, $\lambda$8465) 
and two of VO: ($\lambda$7445, $\lambda$7865). These prominent bands are mainly 
temperature sensitive, making them ideal for the spectral classification.
Figure~\ref{fig:spectra} shows a sequence of K and M-type spectra from the 
WIYN/HYDRA and P200/Norris datasets. 
The TiO absorption bands used to initially classify the stars are indicated. 
Stars earlier than mid-K do not have strong TiO features and atomic
line ratios were used, such as \ion{Ca}{1} $\lambda$6162 vs. \ion{Na}{1} $\lambda$5892 
for F2-K7 stars, \ion{Ca}{1} $\lambda$6162 vs. \ion{Mg}{1} $\lambda$5175,
upper Paschen lines, and other \ion{Ca}{1} and \ion{O}{1} lines for A0-F0 stars. 
Spectral typing for stars earlier than late-K is best done blueward
of our spectral range.

Classifying our red spectra quantitatively 
first involved two dimensional plots of spectral feature index pairs 
such as: TiO $\lambda$6760 vs TiO $\lambda$7100, 
TiO $\lambda$7100 vs TiO $\lambda$7800, 
and  VO $\lambda$7445 vs VO $\lambda$7865.  
These diagrams were used to initially bin the late type 
ONC stars by finding the spectral type along the standard star sequence 
that is closest to the position of the target star.  Second, cubic equations 
were used to fit the standard sequence for each TiO band.  
The standard stars included those at
medium spectral resolution from \citet{1995AJ....109.1379A} 
and \citet{1991ApJS...77..417K} extending up to M6 (near the brown dwarf limit)
and \citet{1999ApJ...519..834K} beyond this.

For the majority of stars, different feature strength ratios predicted
consistent spectral types.  However, using line or band indices gives only
an estimate for the spectral type.  Hence, after assigning preliminary spectral types 
by the quantitative analysis techniques described above, 
each spectrum was examined visually 
against a grid of the standard star spectra in order to verify and in some 
cases slightly modify the quantitative spectral types.  This step was 
undertaken separately by two of the authors.  

Given the age estimate of the ONC, the stars were assumed to be dwarfs 
in the above quantitative classification methods,  
with surface gravity not explicitly taken into 
account for the initial classification.  
For spectral types later than about M5, however, there is a
strong gravity dependence that must be considered in order to avoid 
classifying stars later in type than they really are.  
Since the TiO and VO bands acquire a distinct shape 
in proceeding from dwarf-like to giant-like surface
gravities, particularly the red end of the TiO $\lambda$8465 band,   
this could be accounted for during the visual examination part of the process. 
Surface gravity effects are also seen in the strength of the Na I doublet 
($\lambda\lambda$ 8183, 8194 \AA) at spectral types later than M2.

\begin{figure}
\epsscale{1}
\plotone{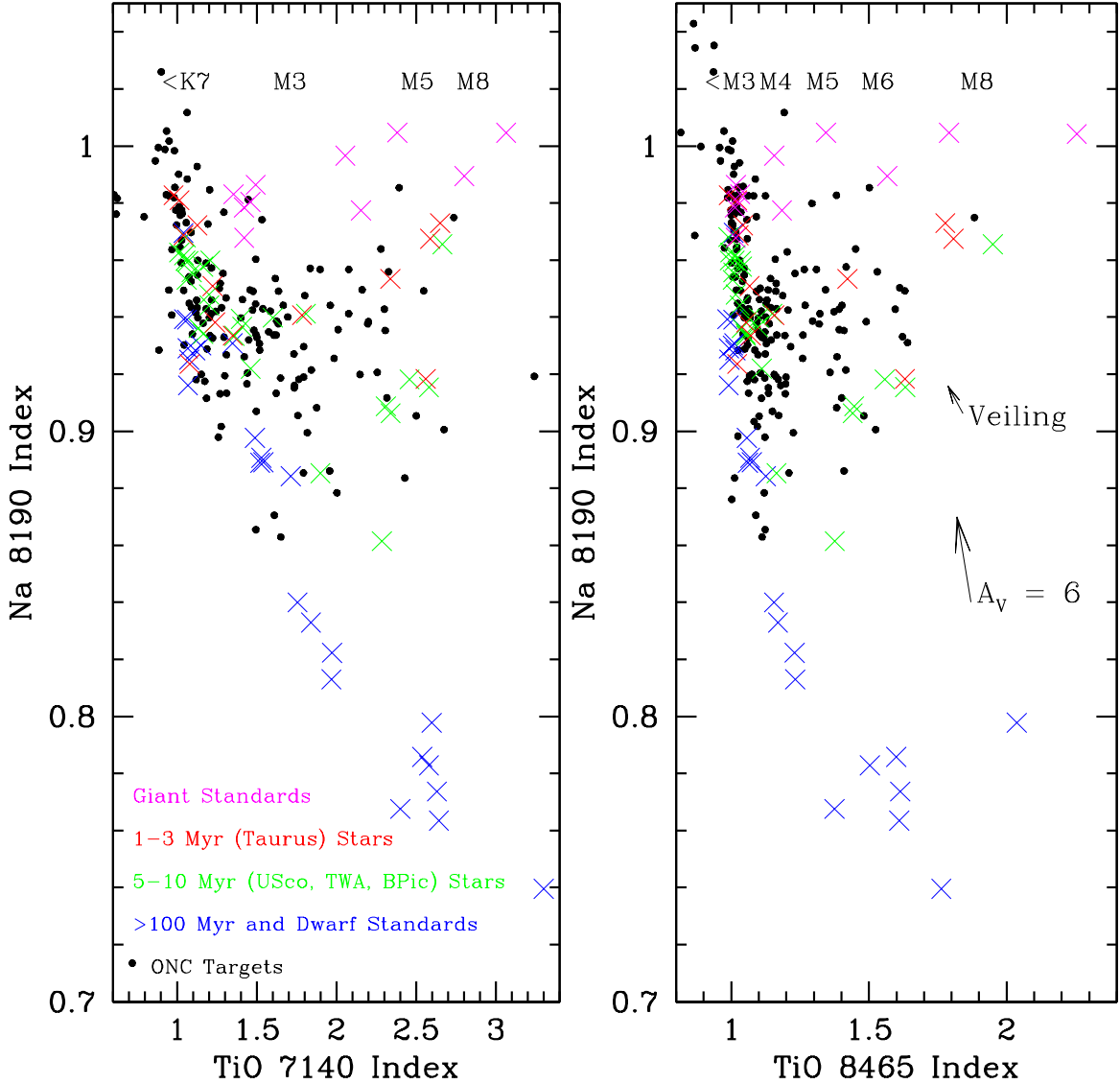}
\caption{The gravity sensitive Na I 8190 \AA\ doublet spectral index 
is plotted as a function of the temperature sensitive TiO 7140 \AA\ index
(left panel) and TiO 8465 \AA\ index (right panel), following  
Slesnick et al. (2006ab).  ONC data (black points)
can be compared to standards of known surface gravity including
giants (magenta crosses), 
dwarfs in the field, Hyades, Pleiades, and AB Dor groups (blue crosses), 
and intermediate-gravity pre-main sequence stars in nearby young 
associations 1-10 Myr old (red and green crosses), 
as indicated in the figure legend.  A signal-to-noise cut has been applied 
to the entire ONC spectral data set with only those objects in the best
20\% of our spectral data shown.  In the right panel, 
the \citet{2006AJ....132.2665S} vectors indicating 
the effects of veiling and reddening are shown.  For the left panel
the veiling vector would be significantly flatter 
\citep{2006AJ....131.3016S} and the reddening vector 
would be only slightly flatter. 
\label{fig:grav}}
\end{figure}

Indeed, being candidate pre-main sequence stars, our objects are expected to
have surface gravities in between those of dwarf stars and giant stars,
though closer to dwarfs.  Thus following \citet{2006AJ....131.3016S}
who were themselves guided by the findings of Kirkpatrick et al. (1991), 
we investigated quantitatively the strength of the
Na I $\lambda$8190 doublet in our spectra. 
This feature weakens towards lower values of surface gravity (see Figure 7 of 
\cite{2008ApJ...689.1295K} for a demonstration). 
We show in Figure~\ref{fig:grav} the same plot 
utilized by  \citet{2006AJ....132.2665S} -- their Figure 4 -- of the 
gravity-sensitive Na I $\lambda$8190 index vs the temperature-sensitive
TiO $\lambda$8465 and TiO $\lambda$7140 indices.  
ONC stars later than spectral type $\sim$M2 typically  
exhibit the intermediate Na I $\lambda$8190 line strengths expected from
young 1-2 Myr old pre-main sequence stars.  
Based on this Figure, we conclude that the vast majority of 
the stars M3 and later in our sample can be presumed ONC members 
given the consistency of their surface gravities with the 
assumed young age of the cluster.  
While a few of the M dwarfs could be interlopers along the line of sight
with higher surface gravity,  the ONC sample at I$\la$17.5 mag
appears to be largely uncontaminated by foreground dwarf stars.

For those late type stars with medium to high signal-to-noise ratio 
and that are not severely affected by nebular contamination (which can contribute
strong lines in either the on-feature or off-feature band regions)
or strong accretion (which can render observed spectra too blue relative
to the expected slope for the spectral type), the quantitative method 
is reliable and robust to visual examination.  For stars 
with a significant amount of reddening, low signal-to-noise ratio, or both, 
experimentation has shown that the quantitative technique is in fact more trustworthy 
than ab initio examination by eye.  Indeed, for most early-to-mid M type stars, 
the spectral types adopted were generally those assigned quantitatively.  
For the late M types, by-eye examination was more rigorous with detailed
attention to surface gravity effects needed for accurate classification.  
At earlier types, some adjustments to the quantitative types were made
for nearly all FGK stars.  These were 
at the 1-3 sub-class level for late K stars but more for the earlier types
for which our red region spectra include narrow features that are only weakly 
varying with temperature, and therefore challenging for our coarse
quantitative technique.  Methods employing equivalent width measurement 
would be more suitable for quantitative analysis than our wide bandpass methods
at these earlier spectral types.
As reported above, not all of the spectra were classifiable. 

\subsection{Spectral Classification Quality Control}

In order to assess the quality of the spectral typing, we can look at the sub-set of sources
with more than one spectrum obtained and classified.  As a reminder, we have
936 total spectra of 707 unique sources, of which 185 unique sources were
observed more than once.  Of these, only 89 sources have multiply derived
spectral types; the remainder have only one or zero of the multiple spectra 
classifiable (due to either signal-to-noise or nebular contamination issues).
Among the 89 stars, 78 have two classifiable spectra and 10 have three classifiable
spectra.  One source, H3132, had six spectra obtained of which five are classifiable. 
The range in the resulting, independently derived, spectral types has a median value
of just one sub-class with standard deviation less than two sub-classes. 
Of the 89 repeated sources, 52 are typed within one sub-class 
and 71 within two sub-classes.  The outliers with larger discrepancies 
tend to be G or K stars where accurate spectral typing 
is more difficult from red wavelength data.
Notably,  H3132 exhibited a range of only one sub-class 
among the types derived from the five different spectra, 
which were independently classified as M5.5, M6, M5, M5.5, and M5.
It is thus clear that in the presence of good data and minimal complications 
from astrophysical effects, the methods are robust.  

However, a number of
systematics are present in the Orion Nebula Cluster region that can
affect the spectral types.   The main one is the accuracy of the sky/nebula subtraction.
Some of our spectra have a blue SED but also late-type spectral features,
which indicates that some nebular contamination remains.  Although this slope/spectrum
discrepancy can be seen in rapid accretors as well, the Orion Nebula itself
predominantly causes the effect in our data sets.  Several of the multiply observed 
stars with large spectral type discrepancies are cases in which one spectrum
has much better sky subtraction than the other.  For these stars it is easy to
see that the later spectral type is the more appropriate one; however, other
cases may exist where we have only one spectrum and have classified a star 
too early, e.g. G or K type, when it is really dominated by nebular continuum 
that if accurately subtracted would yield a later spectral type.   
As noted above, the presence of a strong accretion continuum can similarly 
make a spectrum appear blue as well as fill in the atomic absorption lines, 
resulting in a spectral classification that is biased earlier than the true type
and we may indeed have some of these cases.  
Another complication to accurate spectral typing is reddening.  
While we believe our methods are able to account for
modest amounts of reddening, it is possible that in extreme cases 
there are systematic effects that enter and cause larger than average
spectral type errors.   A final worry is the possible presence of cool spots,
similar to those seen on older non-accreting but still active young stars.
These could either produce or enhance late type features on the red side of our
spectral range, leading to a bias towards a later spectral type assignment
than truly characterizes the underlying star.  In summary, one or more of the
above astrophysical effects could be present and therefore there are possible
spectral type biases in both directions.  

Finally, we note that for comparison purposes we also ran our spectra through
the distributed ``SpTclass" software developed by Hernandez (2005).  
We found that 
the combination of spectral resolution and signal-to-noise effects, plus
perhaps the lack of flux calibration in our dataset,
rendered the SpTclass results suspect.  Specifically, the results are
somewhat degenerate for stars we classified by hand as F through K7 stars, 
with little useful output.  Even for the standard
stars in this spectral type range there was quite a lot of scatter and
a bias in the F, G, and early K range for SpTclass to classify stars 
much later than their true types; in the K5-K8 range, however,
there was good agreement.  For the later type objects, M0-M9, we found
a systematic offset in both our object classifications and tests with
standard stars of approximately 2 to 2.5 spectral subtypes, now in 
the sense that the SpTclass results were too early compared to the actual 
spectral type.
We believe that this comparison demonstrates the need to classify stars 
relative to a grid of standards taken with identical or 
similar equipment.  Furthermore, as our spectra are not flux-calibrated 
having been taken with fibers rather than slits, there may be an additional 
explanation for the systematic effects reported above if the assumptions 
of the SpTclass software regarding continuum shape are being violated in our dataset.  
Regardless of any specific SpTclass biases, 
our experience suggests general caution when considering
automated techniques for spectral classification, 
though they may be very useful when applied
to data sets consistent with their development or in less
challenging spectroscopy environments than the ONC.

\subsection{Spectral Type Results}
\begin{figure}
\includegraphics[scale = .9]{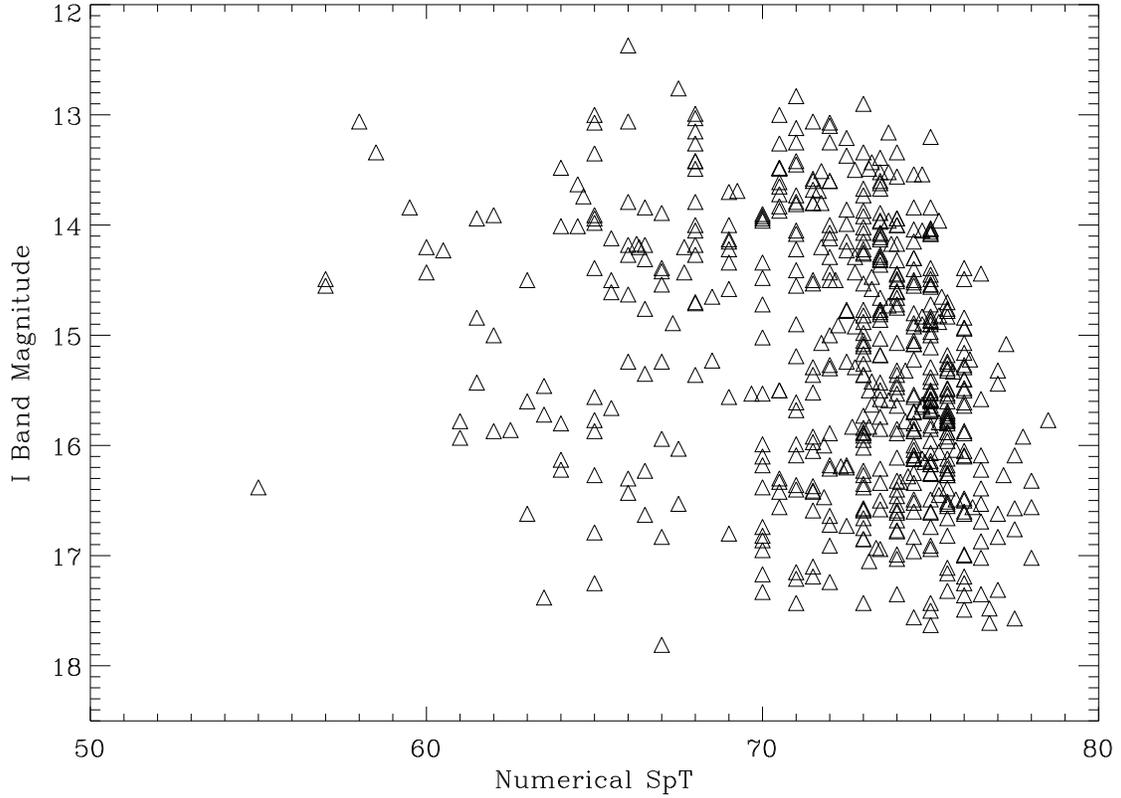}
\caption{
I-band magnitude from H97 vs spectral type from Table 2.  
Along the abscissa, spectral type (SpT) is represented numerically
in a scheme where 50 corresponds to G0, 60 to K0, and 70 to M0.
There is a spread in
the apparent magnitudes of $\sim$5$^m$ that is roughly constant 
with spectral type. We attribute the scatter 
largely to differential reddening, though variations
among stars of the same spectral type in the amount of 
accretion-related continuum excess emission or in intrinsic radii 
are also possible contributors. 
\label{fig:ivsspt}}
\end{figure}

The final spectral types for our ONC sample are presented in Table 2, where 555
are from the new WIYN/Hydra data and 99 are from the new Palomar/Norris data, 
with 35 stars classifiable in both spectral data sets.  
Our newly derived spectral types are on the same system as those in
H97, having been assessed by the same author.  
Figure~\ref{fig:ivsspt} shows the run of apparent I-band magnitude with
spectral type.  The scatter at any given magnitude can be attributed to
cumulative differences in foreground extinction, accretion continuum excess, 
and/or stellar radius.  For 254 of the 619 stars, 
the spectral types are the first 
ever reported. An additional 63 stars have spectral types in other literature 
more recent than the H97 study and compilation, as cited in Table 2.

\subsection{CaII Emission Line Results}

The strong continuum and line emission from the Orion Nebula combined with 
the difficulties of sky subtraction
using multi-fiber instruments in high background fields such as the ONC,
means that many of the emission lines traditionally used to diagnose 
accretion and outflow in T Tauri stars 
(e.g. H$\alpha$ and other Balmer lines, [NII], [SII] and other forbidden lines) 
can not be used to identify such objects of interest among
our sample stars.   However, \ion{Ca}{2} triplet 
($\lambda\lambda$ 8498\AA, 8542\AA, and 8662\AA )
emission was identified by \citet{1998AJ....116.1816H} 
as being a robust accretion indicator in the ONC,
even in the presence of strong nebular background since 
these high density \ion{Ca}{2} lines typically 
are not seen in nebular spectra at low spectral resolution.  
With the line strengths and profile widths 
that are observed in young stars, \ion{Ca}{2} triplet emission 
is attributed to accretion processes
rather than to chromosphere activity that would produce weaker lines 
\citep{1980ApJ...242..628H, 1992ApJS...82..247H, 1993ApJS...85..315S,
2006A&A...456..225A, 2011MNRAS.411.2383K}.  
The physical conditions responsible for the emission
appear to require high density, $n_H \sim 10^{12}$ cm$^{-3}$
and moderate to low temperature, $T < 7500$ K
\citep{2011MNRAS.411.2383K}.  

\begin{figure}
\includegraphics[scale = .7]{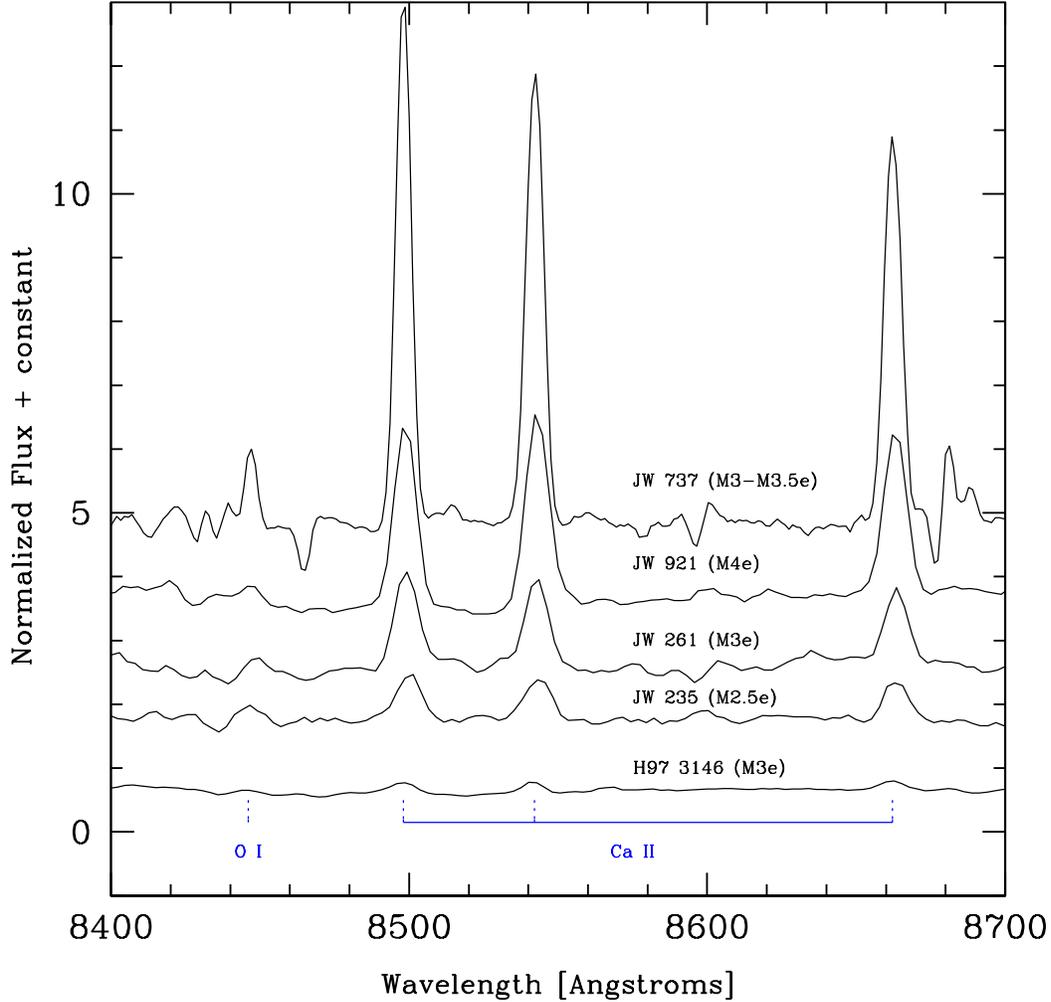}
\caption{
Example \ion{Ca}{2} triplet emission spectra.   The top spectrum
is one of the strongest emission line sources in the sample presented here,
with $\lambda$8542 equivalent width of almost -40\AA, while the bottom spectrum
is one of the weaker emitters at -1.2\AA.  
Note the variation among sources in the triplet line ratios; for classical T Tauri stars 
in Taurus, the $\lambda 8542$ line is generally the strongest with 
typical values of $1.20\pm0.19$ for the 8542 \AA\ to 8498 \AA\ ratio and
$1.28\pm0.13$ for the 8542 \AA\ to 8662 \AA\ ratio
(based on analysis of our own collection of optical spectra for representative samples).  
Most sources show weak \ion{O}{1} emission along with the stronger \ion{Ca}{2}.
\label{fig:caii}}
\end{figure}

\begin{figure}
\includegraphics[scale = .7]{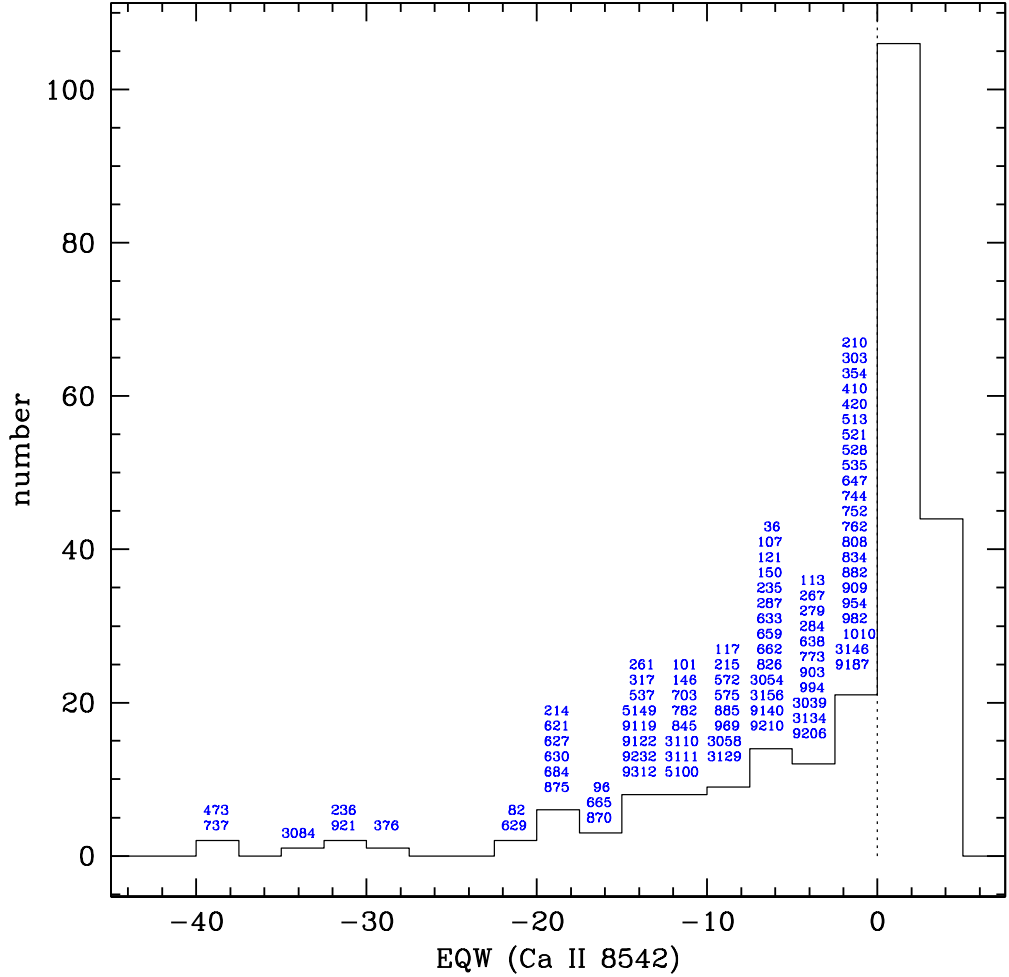}
\caption{
Histogram of \ion{Ca}{2}  $\lambda$8542 equivalent widths (EQW) measured
for several hundred 
of our spectra with adequate signal-to-noise in the continuum.
Labels mark the identifiers from Table 2 of 
the readily diagnosed \ion{Ca}{2} triplet emitters;
each star appears only once, with equivalent widths measure from different
spectra of the same object averaged.
Approximately $\sim$ 40\% of the measured spectra and
$\sim$20\% of the total sample have notable emission. 
Some stars with multiple spectra exhibited line variability 
up to the 25\% level, which could move them between bins on this plot.
\label{fig:caiihist}}
\end{figure}

Approximately 20\% of our stars exhibit \ion{Ca}{2} triplet emission, 
consistent with the fraction reported in the earlier \citet{1998AJ....116.1816H} study.  
Figure~\ref{fig:caii} shows some example spectra of strong, medium, and weak emitters
which occur mainly among the late K and early M spectral types while
Figure \ref{fig:caiihist} shows a histogram of the measured 
\ion{Ca}{2} $\lambda$ 8542 (pseudo) equivalent widths. 
We measured equivalent widths using $splot$ in IRAF for each of the \ion{Ca}{2} triplet lines
against the pseudo-continuum, as well as the nearby O I 8446 \AA\ line. 
Notes were made concerning the strength of the upper Paschen lines
and whether the \ion{Ca}{2} lines have significant \ion{H}{1} contamination
(plausibly from strong accretion, but more likely from incomplete 
nebular subtraction).  
Our identification of emitters has required that the \ion{Ca}{2}
line series either have the 8542\AA\ line stronger than any 8600\AA\ 
pure \ion{H}{1}, or 8498\AA\ stronger than 8542\AA.
Because the local continuum
is significantly depressed for M-type photospheres relative 
to earlier spectral types, there is an expected spectral type 
dependence of the measured equivalent widths for a given 
constant \ion{Ca}{2} line strength. We also note that many of the stronger
emitters are variable among our several spectra, at a level up to 25\%. 

The range of equivalent width values in the ONC fully populates 
the range seen in other star forming regions.  A linear relationship
between \ion{Ca}{2} triplet line flux and total accretion luminosity
is now well-established in the literature 
(e.g. Muzerolle et al. 1998; Herczeg \& Hillenbrand 2008; Dahm 2008).
With knowledge of $M_\ast/R_\ast$, such measurements can lead to
derivation of mass accretion rates.  In the absence of stellar
parameters, however, and with only equivalent widths rather then line fluxes
(which would require either flux-calibrated spectra
or contemporaneously obtained I-band photometry to mitigate 
errors due to variability) we can not derive accretion rates 
here.  We can comment though, that the previously established
correlations suggest that \ion{Ca}{2} $\lambda$ 8542 equivalent widths
in the range 0.5-40 \AA\ for low mass stars
correspond to accretion luminosities in the range $10^{-5}$ to 0.1 $L_\odot$
and accretion rates from $<10^{-10}$ to $10^{-6.5}$ $M_\odot$/yr.

We conclude that a wide range of accretion rates
characterizes the young ONC population.
Once stellar parameters are better established for our sample,
these several hundred young stars can be further investigated 
for correlations between accretion properties and stellar mass and age.

\section{Comparison with Previous Efforts}

All spectral classification results 
that have been derived from previous spectroscopy
for the H97 optical sample 
are presented in Table 2, along with the spectral types newly derived in this paper. 
The historical data is mostly from the H97 compilation, but also included
are our own revisions and updates based on that same spectral data set\footnote{Some of 
the latest type stars were reclassified either as described in 
\citet{2004ApJ...610.1045S} or here, based on improved attention to the effects 
of surface gravity on temperature diagnostics; these are somewhat subtle 
at low signal-to-noise ratio and were not fully appreciated 
at the time of the H97 study.}
as well as more recent estimates of spectral types by e.g. 
Luhman \& Rieke (2000), Lucas et al. (2001), 
Rhode et al (2001), Wolff et al. (2004), Slesnick et al. (2004), 
K. Stassun (private communication, 2005), 
Riddick et al. (2007), H.C. Stempels (private communication, 2009), 
Weights et al. (2009), Daemgen et al. (2012), and Correia et al. (2013) as referenced.
We note that for the Weights et al. (2009), Riddick et al. (2007), and Lucas et al. (2001)
papers it is not straight forward to associate the rounded or truncated
coordinate naming system of the authors with objects in our list of interest, especially
as the positions for nominally the same objects have changed for some
sources between these various studies.
We have done our best to match the reported spectral types for these low precision
coordinate-based names to known objects within 2", where possible based on
similarity of reported apparent magnitudes.  

Spectral types from the sources listed above for stars
that are fainter than the compiled H97 sample appear in Table 3.  Note that unlike
Table 2, the astrometry reported in Table 3 has varying precision.  To the best of our
knowledge, Tables 2 and 3 represent the complete list of spectroscopically determined
spectral types for ONC stars.   Additional photometric spectral types may be found
in \citet{DaRio10}. 

Considering now only Table 2, of the 1576 optically visible stars 
cataloged in H97, about 500 lacked a spectral type 
or any spectral type constraint at that time,
while now just under 200 lack a spectral type estimate.
Several tens of stars have only wide ranges or limits for their 
estimated types, that is, just constraints on the latest or the
earliest type that is consistent with the observed spectrum.
Many of the stars lacking adequate classification  are in regions of 
bright nebulosity such that repeated observations at optical wavelengths
have failed to yield a classifiable spectrum.  Others are close companions to
brighter stars and likewise have not had suitable spectra obtained.

\subsection{Comparison to Hillenbrand (1997) Spectroscopic Types}

As reported above, 619 stars are typed from our Norris and HYDRA data.  
While the observational focus 
of the new data acquisition was on those stars without previously determined 
spectral types, in practice we placed fibers on other ONC members 
rather than not assigning them at all. 
Thus the set of stars for which spectral types are derived in this paper 
overlaps significantly with those in the H97 paper.  
This enables comparison of results
obtained with different equipment but using similar analysis methods

\begin{figure}
\includegraphics[scale = .9]{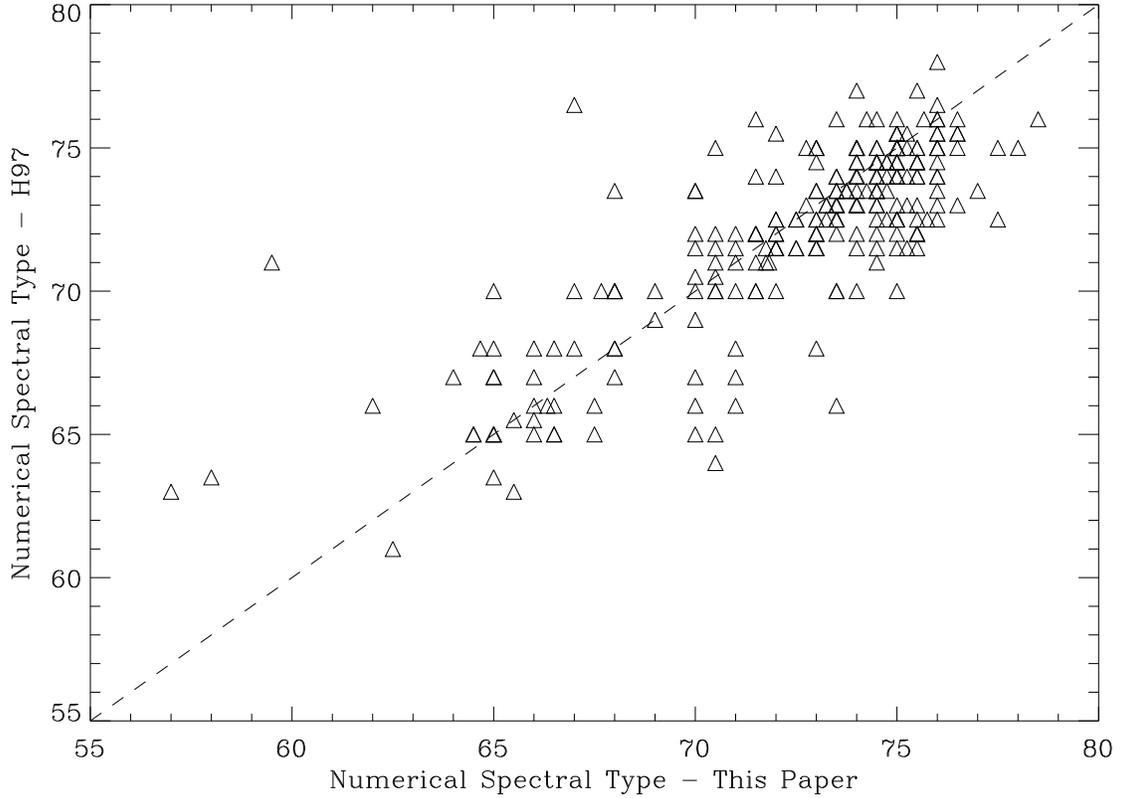}
\caption{
Comparison of spectral types quoted in H97
with those quoted in the last column of Table 2,
derived in this study. 
Along both axes, spectral type is represented numerically
in a scheme where 50 corresponds to G0, 60 to K0, and 70 to M0.
The dashed line represents one-to-one correspondence with
the root-mean-squared deviation 2.25 spectral sub-classes. 
The largest discrepancies can be attributed to obvious errors
in sky subtraction in one or the other set of spectra;
this effect probably contributes to the lower level scatter as well. 
\label{fig:compare_h97}}
\end{figure}

In the vast majority of cases 
there is remarkable agreement between the previously and presently reported 
spectral types, as can be seen from examination of Table 2. 
This is despite the known observational challenges and the 
likely influence of spectrophotometric variability due to real 
changes in accretion / extinction parameters.
Figure~\ref{fig:compare_h97} illustrates the direct correlation
between H97 and this work; the distribution
exhibits a root-mean-squared scatter of 2.25 spectral sub-classes.
However, some large discrepancies exist as well, notably 
when early-K vs mid-M spectral types have been derived for the same star.   
We advise in these cases that the later spectral type is likely 
the more accurate one since, as discussed earlier,
there are plausible ways to make spectral types appear much earlier 
than the true type in some data.

\begin{figure}
\includegraphics[scale = .5]{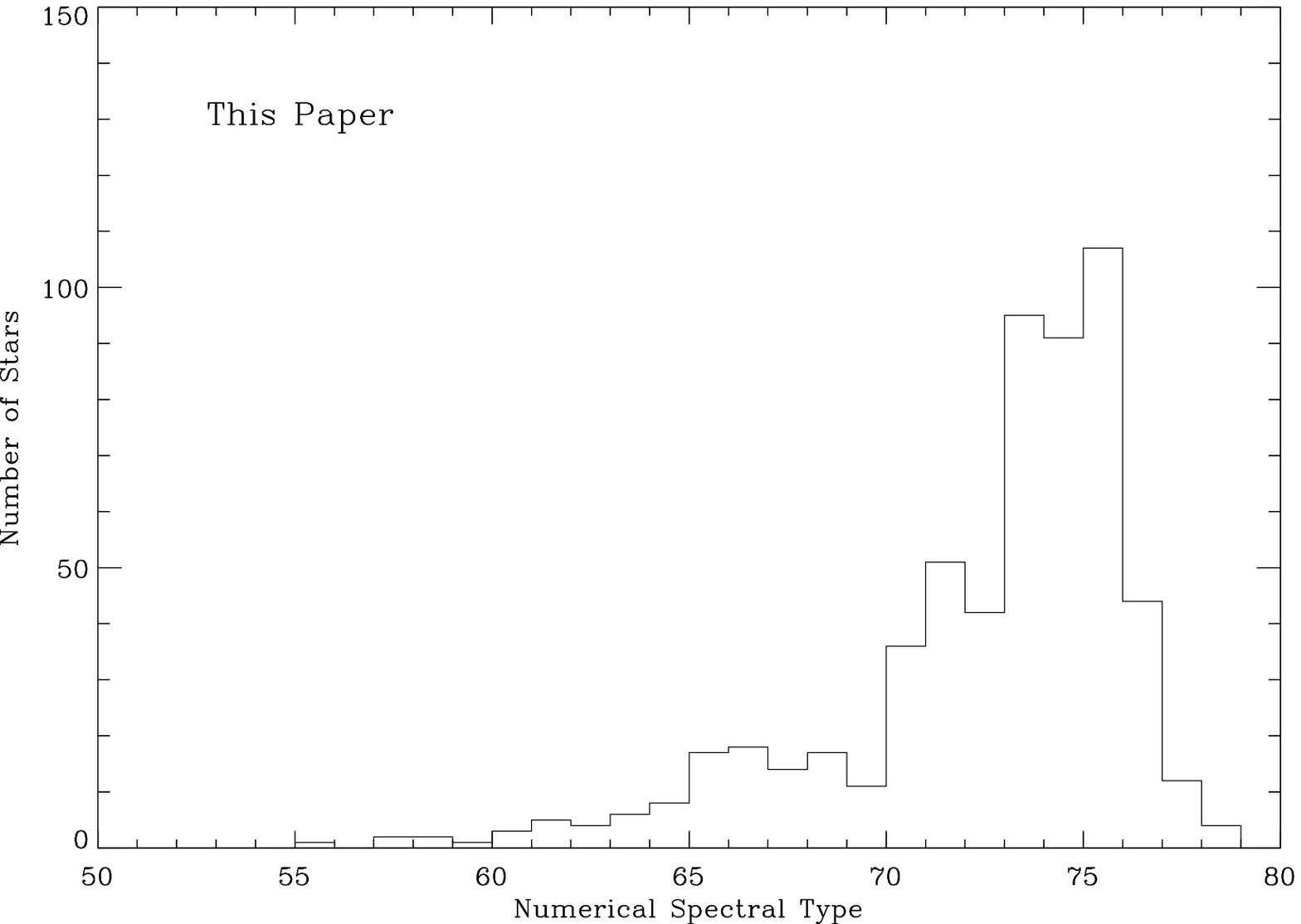}
\\
\\
\includegraphics[scale = .5]{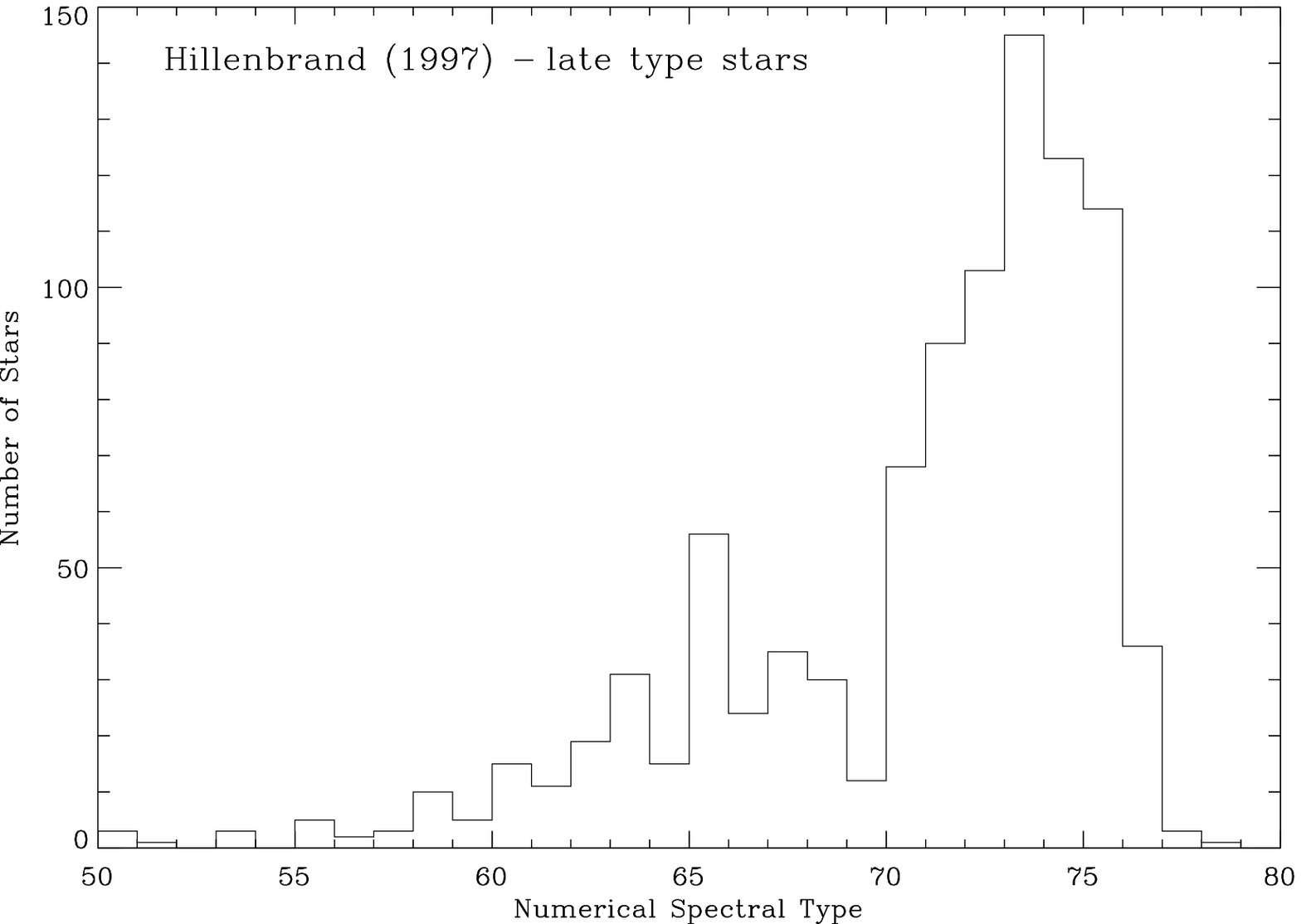}
\caption{Histograms comparing the samples of newly derived spectral types
and those previously reported in H97. 
Along the abscissa, spectral type is represented numerically
in a scheme where 50 corresponds to G0, 60 to K0, and 70 to M0.
\label{fig:spthist}}
\end{figure}

Histograms in Figure~\ref{fig:spthist} show the distribution 
of stellar types from H97 compared to the distribution in the 
new spectral dataset. While the histograms are similar in terms of rising
from earlier to later types with a peak around M5, 
there are notably fewer extremely late type stars seen here compared to 
what was reported in the H97 paper.

\subsection{Comparision to da Rio (2010, 2012) Photometric Types}

\citet{DaRio10} derive spectral types for 217 late type stars in the ONC 
based on narrow band photometry, employing a novel TiO 
photometric index defined as the measured magnitude in a 200 \AA\ 
wide filter centered at 6200 \AA\ relative to the magnitude at
that same wavelength interpolated from a linear fit to V- and I-band
photometry.
\citet{DaRio12} obtained a deeper photometric dataset and used a similar 
technique based on a longer wavelength 7700 \AA\ narrow band filter,
to derive spectral types and extinction values for 1280 objects.
The TiO index (narrow band photometric) technique is in fact similar 
in spirit to the spectroscopic TiO band indices that we have employed here.
The photometric spectral types from these two da Rio studies 
range between M2 and M6, where the technique is valid, 
and have estimated errors of typically 1-2 sub-classes. 

\begin{figure}
\includegraphics[scale = .9]{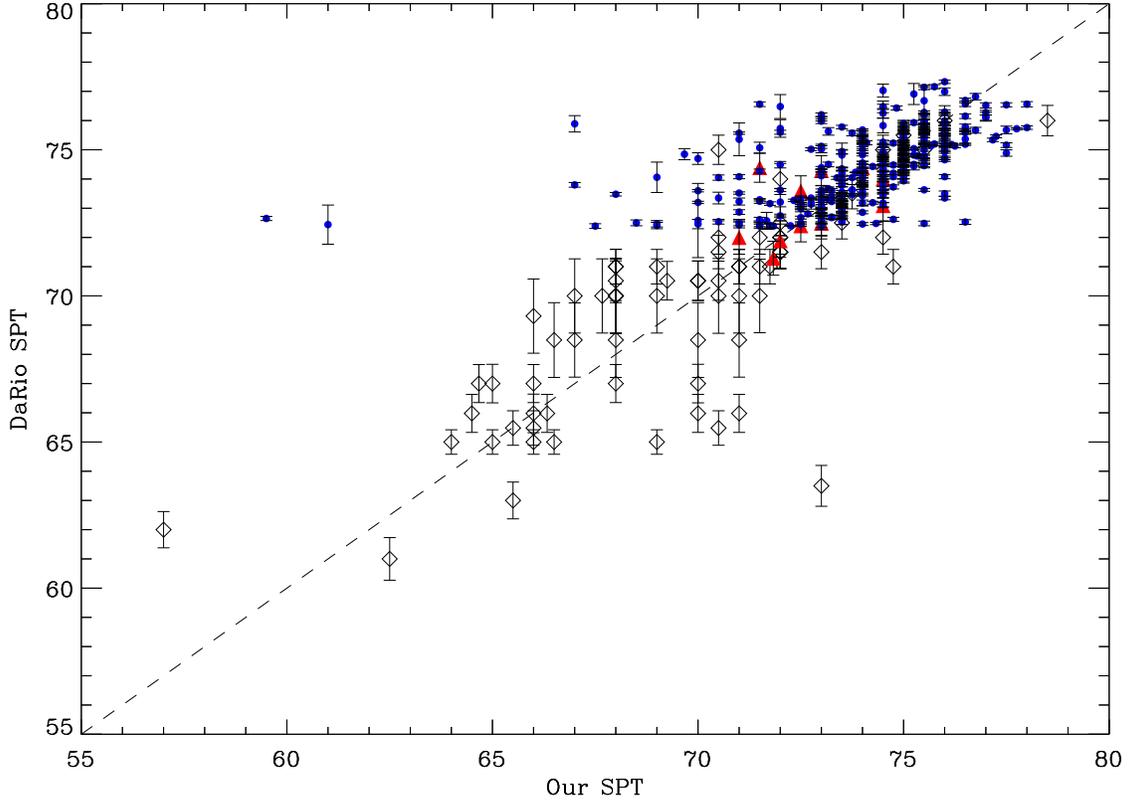}
\caption{
Comparison of spectral types (SpT) quoted in \citet{DaRio12}
with the ones quoted in the last column of Table 2, derived in this study.
Along both axes, spectral type is represented numerically
in a scheme where 50 corresponds to G0, 60 to K0, and 70 to M0.
Filled circles represent spectral types based on the 7700 \AA\ TiO index 
developed by \citet{DaRio12}. Filled triangles
are sources without 7700 \AA\ types but instead have their spectral types
derived from a similar 6200 \AA\ TiO index in \citet{DaRio10}.
Open diamonds do not have photometric TiO types available and so the comparison
for these generally earlier type stars is to the spectroscopically determined 
spectral type quoted in H97, which was adopted by  \citet{DaRio12}.
The dashed line represents one-to-one correspondence with
the root-mean-squared deviation 1.75 spectral sub-classes. 
\label{fig:compare_dario}}
\end{figure}

In Figure~\ref{fig:compare_dario} we compare directly the da Rio 
photometric spectral types with the spectroscopic spectral types
quoted in Table 2; the correlation
exhibits a root-mean-squared scatter of 1.75 spectral sub-classes.
In total, a sample of 437 stars is available 
for comparison between the spectroscopic and photometric spectral types,
with all 437 of those having 7700 \AA\ TiO spectral types, and 315
having 6200 \AA\ TiO spectral types as well, when both photometric
methods could be applied.  While the scatter seems large between the photometric
and spectroscopic methods (filled symbols), it appears to be no worse 
at these late types than
between the new spectroscopy and the old spectroscopy (open symbols)
probing a slightly earlier spectral type range.  It should be noted that,
as discussed above, the error bars in this earlier regime are larger 
than those suggested at the later types from either the photometric 
methods (1-2 sub-classes) or the spectroscopic methods 
(0.5 sub-classes based on repeated robust measurements of the same star 
but up to 2 sub-classes including systematic effects).

\citet{DaRio12} also devised procedures to de-redden their photometry
and obtain both A$_V$ and T$_{eff}$ for individual late type stars.
We have mimicked their analysis by performing synthetic photometry on
our spectra based on the bandpasses specified by \citet{DaRio12}.  Although the
spectra we have are not flux calibrated, because the TiO index on-band
and off-band filter profiles are so close in wavelength this should have only
a minor effect.  The resulting color-color diagram 
based on our spectra (not shown) is indeed
similar in appearance to the one based on narrow band photometry
that is presented in \citet{DaRio12} -- their Figures 5 and 11. 
Specifically, there is a narrow finger of stars having no indication 
of TiO depression but a range of reddening values, and then a spread along 
the TiO index or temperature-sensitive axis,
as well as along a nearly orthogonal reddening vector.

We have further followed the \citet{DaRio12} methodology for deriving
both A$_V$ and T$_{eff}$, finding for our sample extinction values 
in the 0-3 magnitude range and effective temperatures in the 
2700 to 3900 K indicated.
While illustrative of the consistency between different methods for
deriving spectral types and then physical parameters from them,
it should be noted that the scatter between
the two TiO photometric methods (TiO 6200 \AA\ in \citet{DaRio10} vs
TiO 7700 \AA\ in \citet{DaRio12}) is itself 0.03 in log $T_{eff}$,
corresponding to 1.5 spectral sub-classes in the M star regime, and
comparable to the estimated errors in the photometric spectral typing method.

We conclude for mid-to-late M-type stars,
that spectral typing through the use of
narrow-band photometric methods is no less accurate 
in the mean than spectral typing through spectroscopic methods.
For early M stars, however, there is a systematic discrepancy
with later types being assigned through the photometric method than
appear reasonable from the spectroscopy.  This could be due to
nebular contamination in the spectra, the photometry, or both.

\subsection{Spectral Type Variables}

There is good correlation demonstrated overall in Figures 
~\ref{fig:compare_h97} and ~\ref{fig:compare_dario}, as well as
general agreement over many decades in the spectral types
derived by different authors for the same stars
(see Table 2).  However, there are some cases 
where large discrepancies are apparent.

\begin{figure}
\includegraphics[scale = .8]{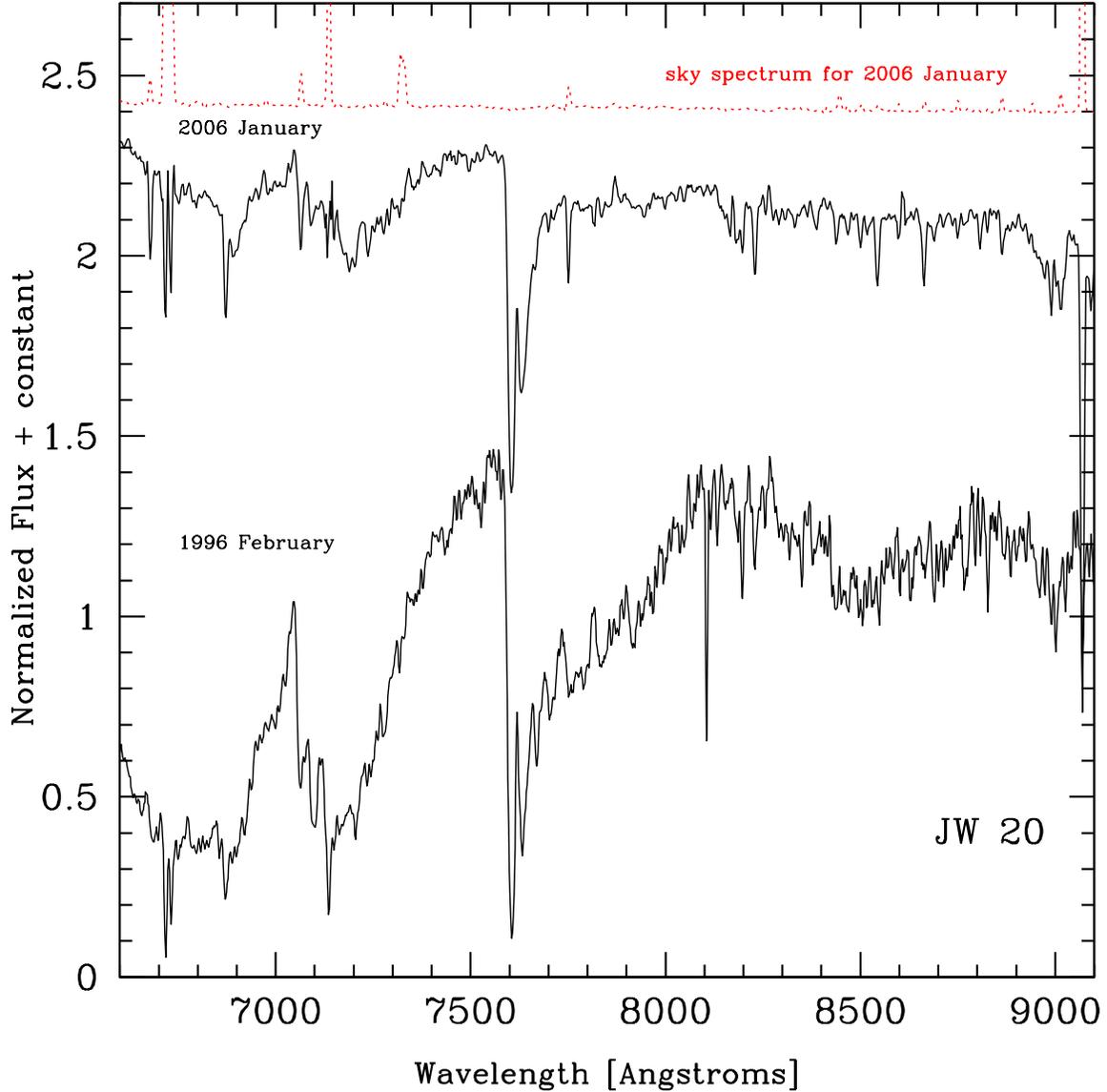}
\caption{
Example of an object, JW 20, whose spectrum appears to have varied
significantly between observations taken about ten years apart.
There is no apparent instrumental or observational reason for
the discrepancy and an astrophysical origin is suggested.
Furthermore, while a large change in accretion-induced veiling 
could change the continuum and weaken the TiO absorption, it is hard
to explain the concomitant appearance of \ion{Ca}{2} triplet absorption.
A scaled down sky spectrum is also shown, to illustrate the spectral regions
where nebular line contamination are most prominent.
\label{fig:jw20}}
\end{figure}

We highlight the case of JW 20 as an illustration.
The H97 spectral type (based on data from 1996, February) 
was M3.5. However, the spectrum presented here 
(based on data from 2006, January) is that of a K8 star,
as illustrated in Figure ~\ref{fig:jw20}.  The star is only
a small-amplitude optical variable \citep[0.05 mag;][]{2002A&A...396..513H} 
with all available literature and catalog near-infrared photometry
consistent within 0.1 mag and the mid-infrared variability
similarly small amplitude \citep{2011ApJ...733...50M}. 
Nevertheless, between the two spectra taken ten years apart,
the deep TiO and VO bands decreased in strength and 
\ion{Ca}{2} triplet absorption was revealed.  

We can not 
explain this discrepancy in terms of signal-to-noise challenges.
Nor can we rationalize an explanation through observational confusion 
since there are no other optical or near-infrared sources within 30".
Data for both spectra were taken with the WIYN/HYDRA 
fiber positioner and there are no indications among the other objects 
of problems with the pointing.  Although the exact position of the sky 
fiber location was different, the sky subtraction method employed was the same,
and we can not see a way of producing these two different
spectra if sky/nebula subtraction accuracy was the only variation.

Careful examination of Table 2 reveals several other such 
stars with large spectral type discrepancies.  We are
driven to the conclusion that in at least some cases, e.g. JW 20 and JW 798,
astrophysical origin must be considered plausible.  In other cases
there may be explanations;  as examples,
for JW 625, H97 3030, and likely others, there is strong 
nebular contamination either on-source or in the available sky positions, 
while for H97 3067 a nearby brighter source could contaminate 
an on-source fiber spectrum during poor seeing conditions. 
These difficulties inherent to crowded, nebular star forming regions
are non-astrophysical sources of discrepant spectral types
that should be considered, in addition to low signal-to-noise, as 
possible classification biases.

\section{Summary}
We have presented an analysis of new low resolution optical spectra
that extends and updates the catalog of spectral types
for optically visible stars in the ONC.  For 254 of our 619 newly typed 
stars, we present the first published spectral classifications.
Of the approximately 1600 sources in the optical photometric catalog 
presented by H97, 90\% now have spectral types or spectral type constraints 
based on low resolution optical spectroscopy, 
compared to just 68\% at the time of that paper and 5\% previous 
to that paper.  Many of the optical stars still have only wide ranges 
or limits on their spectral types, and not single valued spectral types.  
A similar number of optically invisible
stars await the application of infrared spectral typing techniqes, which
will be no less challenging given the complex background of the
Orion Nebula.  

The newly derived and the older optical spectral type 
distributions appear to have similar shape.  
In a point-by-point comparison, spectral classification differences 
are typically around 2 spectral sub-classes. 
We have also compared our grating spectrograph methods to narrow band 
photometry methods for deriving spectral types, and also found that the scatter
of nearly 2 spectral sub-classes is somewhat larger than 
the precision estimates claimed for each method. 
Some young stars may exhibit real spectral variability over time.

With the current set of spectral types as well as recently available 
high quality optical photometry \citep[e.g.][]{DaRio09} 
plus a substantial recent revision in the accepted distance to the ONC, 
it is timely to revisit not only the HR diagram and its derivatives, 
as in \citet{DaRio12}, but also investigations of 
stellar activity, stellar rotation, and circumstellar
disk properties in this star forming region.

\acknowledgments
We acknowledge with gratitude the expert assistance of Todd Small 
concerning usage of the Palomar/Norris spectrograph and its data products,
as well as team HYDRA at KPNO/WIYN. 
Emma Hovanec helped with the measurement and tabulation of emission lines.  
We thank Nicola da Rio for providing material
that enabled the comparison shown in Figure 10, 
as well as Keivan Stassun and Eric Stempels for communication of 
unpublished spectral type information.
Some financing for the middle stages of this work was provided 
in the mid-2000's through the HST/Treasury program on the ONC (PI Massimo Robberto).
This research has made use of the SIMBAD database,
operated at CDS, Strasbourg, France.

{\it Facilities:} \facility{Hale (Norris)}, \facility{WIYN (HYDRA)}.

\clearpage

{\small
\begin{deluxetable}{cccccccc}
\tabletypesize{\footnotesize}
\tablecaption{Observation Log}
\tablehead{\colhead{Configuration} & \colhead{Date} &\colhead{Instr.\tablenotemark{a}\tablenotemark{b}} & \colhead{On-Targets } & \colhead{Offset Sky } & \colhead{\# Target } &  \colhead{\# Sky } & \colhead{Comments} \\
\colhead{Name } & \colhead{} &\colhead{} & \colhead{T$_{exp.}$ (sec)} & \colhead{T$_{exp.}$ (sec)} & \colhead{Fibers\tablenotemark{c}}&  \colhead{Fibers\tablenotemark{c}}& \colhead{} \\
} 
\startdata
Nor    & 5 Sep. 1999 & Norris& 2700 & 900 & 120 & 13 & \\
f1d  & 13 Jan. 2006 & Hydra & 5400 & 2700\tablenotemark{d} & 55 & 31& thick clouds\\
f2c  & 13 Jan. 2006 & Hydra & 5400 & 2700\tablenotemark{d} & 60 & 29& thick clouds\\
b2d  & 15 Jan. 2006 & Hydra & 7200 & 2400\tablenotemark{d} & 30 & 54& thick clouds\\
C1   & 5  Jan. 2007 & Hydra & 1440 & 418 & 61 & 28 \\
C2   & 6  Jan. 2007 & Hydra & 1200 & 600 & 50 & 36\\
M1   & 6  Jan. 2007 & Hydra & 2400 & 900 & 69 & 20 & poor seeing, quality\\
N2   & 6 Jan. 2007  & Hydra & 2200 & 750 & 68 & 20 & poor seeing, quality\\
G1b  & 6 Jan. 2007  & Hydra & 3600 & 2100\tablenotemark{d} & 69 & 19 & poor seeing, quality\\
G2b  & 6 Jan. 2007  & Hydra & 4800 & 1915\tablenotemark{d} & 67 & 21 \\
D3b  & 7 Jan. 2007  & Hydra & 1000 & 500 & 38 & 46\\
N3d  & 7 Jan. 2007  & Hydra & 2000 & 700 & 63 & 27\\
G3b  & 7 Jan. 2007  & Hydra & 4500 & 1500\tablenotemark{d} & 61 & 25\\
Z4b  & 7 Jan. 2007  & Hydra & 2000 & 700 &  70 & 20\\
W5a  & 7 Jan. 2007  & Hydra & 1800 & 600 &  70 & 20\\
W6c  & 7 Jan. 2007  & Hydra & 1100 & 600 &  71 & 19
\enddata
\tablenotetext{a}{ The Norris Spectrograph was used in a setting 
covering $\lambda$ 6100-8750\AA\ at R$\sim2000$.}
\tablenotetext{b}{ The Hydra Spectrograph was used in a setting 
covering $\lambda$ 5000-10000\AA\ at R$\sim1500$.}
\tablenotetext{c}{ Number of fibers assigned to target / sky positions.}
\tablenotetext{d}{ Multiple exposures were combined resulting in the total exposure time.}
\end{deluxetable}
} 
\clearpage

\begin{deluxetable}{cllp{6cm}l}
\tabletypesize{\footnotesize}
\tablecaption{ONC Spectral Types - ABBREVIATED VERSION; SEE JOURNAL FOR FULL ELECTRONIC TABLE}
\tablehead{\colhead{Identifier \tablenotemark{a}} & \colhead{Right Ascension \tablenotemark{b}} &\colhead{Declination \tablenotemark{b}} & \colhead{Previously Reported SpT \tablenotemark{c}} & \colhead{Newly Reported SpT \tablenotemark{d}} } 
\startdata
3132  &  5:34:11.52 &  -5:30:19.7 &  \nodata  &  M5.5    \\ 
3157  &  5:34:11.70 &  -5:33:55.9 &  M1(H)  &  \nodata    \\ 
3126  &  5:34:12.91 &  -5:28:48.1 &  K7(H)  &  \nodata    \\ 
3153  &  5:34:13.07 &  -5:33:48.3 &  M3(H)  &  \nodata    \\ 
3156  &  5:34:13.22 &  -5:33:53.5 &  M2.5e(Sta)  &  M1.5,M2e    \\ 
3035  &  5:34:13.70 &  -5:17:43.8 &  G:(H)  &  K4-M1    \\ 
3118  &  5:34:14.49 &  -5:28:16.6 &  G6(H)K0(WSH)  &  \nodata    \\ 
3007  &  5:34:14.90 &  -5:14:18.0 &  M2.5(H)  &  M3.5    \\ 
3024  &  5:34:14.97 &  -5:15:49.3 &  M1(H)  &  \nodata    \\ 
3078  &  5:34:15.11 &  -5:23:00.0 &  \nodata  &  M2.5      
\enddata
\tablenotetext{a}{ The star numbers are those listed from 
\citet{1988AJ.....95.1755J} between \#1-1053, 
Parenago (1954) between \#1054-2999, 
\citet{1994ApJ...421..517P} for sources given as 9000$+$ the Prosser number,
and \citet{1997AJ....113.1733H} for sources given as 3000, 5000, 6000 series numbers.
}
\tablenotetext{b}{ J2000.}
\tablenotetext{c}{ The sources of the literature spectral types (SpT) are as listed in Hillenbrand (1997) with more recent additions as
detailed in the text.  The codes are as follows:
\\         B      = Blanco, 1963
\\         CK     = Cohen \& Kuhi, 1979
\\         C      = Correia et al., 2013
\\         D      = Duncan, 1993
\\         Dae    = Daemgen et al., 2012
\\         E      = Edwards et al., 1993
\\         GS     = Greenstein \& Struve, 1946
\\         H      = Hillenbrand, 1997 and subsequent updates to electronically available table.
                         An ``$e$'' indicates emission in the CaII triplet lines.
                         A ``$<$'' indicates spectral type is earlier than that
                             listed while ``$>$'' indicates spectral type is later than that listed.
\\         Ham    = C. Hamilton, 1994 unpublished masters thesis
\\         Her    = Herbig, quoted in Walker, 1969
\\         HP     = Herbig, private communication, 1996
\\         HT     = Herbig \& Terndrup, 1986 or reference therein
\\         J      = Johnson, 1965
\\         LA     = Levato \& Abt, 1976 or Abt \& Levato, 1977
\\         LDW    = Lallemand, Duchesne, \& Walker, 1960
\\         LR     = Luhman, Rieke, et al., 2000
\\         Luc01    = Lucas et al., 2001 (by matching given positions to known objects of similar magnitude within 2" and translating quoted log-g=4 Teff values into a SpT using the same Teff-SpT calibration employed in H97, largely from CK)
\\         Luc06  = Lucas et al., 2006
\\         M      = McNamara, 1976
\\         P      = Prosser \& Stauffer, private communication, 1995
\\         Par    = Parenago, 1954
\\         Petal  = Penston, Hunter, \& O'Neill, 1975 or Penston, 1973
\\         R      = Rhode, Herbst, Mathieu 2001 identification of SB2's
\\         RRL    = Riddick, Roche, Lucas 2007 (by matching given positions to known objects of similar magnitude within 2")
\\         S      = Strand, 1958 reference (mostly to Sharpless)
\\         Sam    = A.E. Samuel, 1993 unpublished PhD thesis
\\         SBB    = Smith, Beckers, \& Barden, 1983
\\         SHC    = Slesnick, Hillenbrand, \& Carpenter, 2004.  SHC indiates optical spectral types and SHC-ir infrared spectral types.
\\         Sta    = K. Stassun, private communication, 2005, February; low dispersion spectra
\\         Ste    = H.C. Stempels, private communication, 2009, March; high dispersion spectra    
\\         T      = Trumpler, 1931
\\         vA     = van Altena et al., 1988
\\         W      = Walker, 1983
\\         WLR    = Weights et al., 2009 (by matching given positions to known objects of similar magnitude within 2"; method is similar to Luc and latest types may be too late)
\\         WSH    = Wolff, Strom, \& Hillenbrand, 2004
}
\tablenotetext{d}{ Spectral type derived in the present study, from the Kitt Peak / WIYN
data and/or the Palomar / Norris data. Many stars were classified based
on several different spectra and in the cases of disagreements
multiple types are listed.
}
\tablenotetext{e}{ Several optically identified objects 
from previous studies apparently are plate defects (JW 459, JW 699)
or nebular knots (PSH 9081, PSH 9326, H97 3071, H97 3089) rather than 
true stellar point sources.
We list them here for completeness but these sources should be removed
from future list of ONC stellar objects.
}
\tablenotetext{f}{These sources have nebular contamination in our spectra but may include some \ion{Ca}{2} emission contribution.}
\end{deluxetable}

\clearpage

{\small
\begin{deluxetable}{cllp{6cm}}
\tabletypesize{\footnotesize}
\tablecaption{Additional ONC Spectral Types: Stars Fainter than those in Table 2}
\tablehead{\colhead{Identifier \tablenotemark{a}} & \colhead{Right Ascension \tablenotemark{b}} &\colhead{Declination \tablenotemark{b}} & \colhead{Previously Reported SpT \tablenotemark{c}}  \\} 
\startdata
[H97b] 10306             & 05:34:55.90 & -05:21:08.5 & M6.25(RRL)  \\ \relax
[H97b] 10313             & 05:34:56.95 & -05:21:21.9 & M6.75(RRL)  \\ \relax
[H97b] 20349             & 05:35:00.90 & -05:21:07.3 & M8(WLR)       \\      \relax              [OW94] 013-306   	 & 05:35:01.3  & -05:23:06   & M7(Luc01)M9(WLR)  \\ \relax
[H97b] 10343             & 05:35:01.37 & -05:24:13.3 & M6(Luc01)M6.5(RRL)M7.5(WLR)  \\ \relax
[OW94] 016-319           & 05:35:01.6  & -05:23:19   & L2.5(Luc01)$>$M9.5(WLR)  \\ \relax
[OW94] 016-430           & 05:35:01.6  & -05:24:30   & M9(Luc06)M9(WLR)  \\ \relax
[H97b] 10348             & 05:35:01.87 & -05:23:53.7 & M5(Luc01)M5.75(RRL)M6.5(WLR)  \\ \relax
[H97b] 20339      	 & 05:35:01.80 & -05:21:06.7 & K2(Luc01)    \\  \relax                   [H97b] 10353             & 05:35:02.31 & -05:21:23.4 & $>$M9.5(WLR)  \\ \relax
[OW94] 031-524           & 05:35:03.1  & -05:25:24   & M7.5(RRL)M8(WLR)  \\ \relax
2MASS J05350313-0525364  & 05:35:03.13 & -05:25:36.5 & M8.75(RRL)$>$M9.5(WLR)  \\ \relax
[H97b] 10364             & 05:35:04.19 & -05:20:12.0 & M6.5(Luc01)M7.75(RRL)M8(WLR)  \\ \relax
[H97b] 20296             & 05:35:04.44 & -05:22:19.5 & late-M(RRL)  \\ \relax
2MASS J05350445-0525264  & 05:35:04.46 & -05:25:26.5 & M8.5(WLR)   \\ \relax
[H97b] 20282             & 05:35:04.62 & -05:22:44.8 & late-M(RRL)  \\ \relax
2MASS J05350467-0525508  & 05:35:04.68 & -05:25:50.8 & M8.5(RRL)  \\ \relax
[H97b] 20302             & 05:35:04.95 & -05:21:42.8 & mid-M(RRL)  \\ \relax
2MASS J05350557-0521407  & 05:35:05.57 & -05:21:40.7 & $>$M9.5(WLR)  \\ \relax
[OW94] 057-247           & 05:35:05.7  & -05:22:47   & $>$M9(Luc06)$>$M9.5(WLR)  \\ \relax
[OW94] 061-401           & 05:35:06.10 & -05:24:00.6 & M8(Luc01)$>$M9.5(WLR)  \\ \relax
[HC2000] 509             & 05:35:06.35 & -05:22:11.6 & M2-M5(SHC)M2-M7(SHC-ir)  \\ \relax
[OW94] 066-433           & 05:35:06.6  & -05:24:33   & K4.5(Luc01)  \\ \relax
[H97b] 20270             & 05:35:07.06 & -05:25:00.9 & M7-M9(SHC)  \\ \relax
[H97b] 10380             & 05:35:07.23 & -05:26:38.6 & M6.5(RRL)  \\ \relax
[OW94] 073-205           & 05:35:07.3  & -05:22:05   & M6(Luc01)  \\ \relax
[HC2000] 743             & 05:35:08.10 & -05:23:15.2 & M6:(SHC-ir)  \\ \relax
[HC2000] 433             & 05:35:08.11 & -05:22:37.5 & M8(SHC-ir)  \\ \relax
[HC2000] 400             & 05:35:08.22 & -05:22:53.2 & M7-M8(SHC)M9(SHC-ir)M8.5(RRL)   \\ \relax
[HC2000] 725             & 05:35:08.27 & -05:23:07.8 & M7:(SHC-ir)  \\ \relax
[H97b] 20335             & 05:35:08.31 & -05:19:37.2 & M7.25(Luc01)M9(WLR)     \\     \relax           [OW94] 084-104   	 & 05:35:08.32 & -05:21:02.4 & L0(Luc01)$>$M9.5(WLR)   \\ \relax    [HC2000] 749             & 05:35:08.34 & -05:23:21.9 & M8(SHC-ir)  \\ \relax
[H97b] 20208             & 05:35:08.44 & -05:23:04.9 & M7(RRL)  \\  \relax
[H97b] 10391             & 05:35:08.42 & -05:22:30.3 & M2(SHC-ir)  \\ \relax
[OW94] 086-324           & 05:35:08.62 & -05:23:24.4 & M5.5:(SHC-ir)   \\ \relax              2MASS J05350865-0520223  & 05:35:08.66 & -05:20:22.4 & M9.5(WLR)  \\ \relax
[HC2000] 455             & 05:35:08.93 & -05:22:30.0 & M2-M6(SHC-ir)  \\ \relax
[HC2000] 724             & 05:35:09.03 & -05:23:26.3 & M6:(SHC-ir)  \\   \relax
[OW94] 092-532           & 05:35:09.2  & -05:25:32   & M7.5(Luc06)M7.5(WLR)  \\ \relax               [H97b] 20298             & 05:35:09.20 & -05:26:05.5 & M8(RRL)  \\ \relax
[H97b] 10403             & 05:35:09.57 & -05:19:42.7 & M6.5(Luc01)M6.75(RRL)M9(WLR)  \\  \relax
[HC2000] 722             & 05:35:09.79 & -05:24:06.7 & M6.5(SHC)  \\  \relax
[H97b] 20184             & 05:35:09.91 & -05:24:10.5 & M3(SHC)  \\  \relax
[HC2000] 62              & 05:35:10.03 & -05:25:01.5 & M9(SHC-ir)  \\  \relax
[HC2000] 90              & 05:35:10.38 & -05:24:51.6 & M7.5(SHC-ir)  \\  \relax
[OW94] 107-453           & 05:35:10.7  & -05:24:53   & M8(Luc06)M8(WLR)  \\  \relax   
[HC2000] 529             & 05:35:10.88 & -05:22:06.0 & M8(SHC-ir)  \\  \relax
[HC2000] 111             & 05:35:11.15 & -05:24:36.5 & M9(SHC-ir)  \\  \relax
[HC2000] 434             & 05:35:11.20 & -05:22:37.8 & M2(SHC-ir)  \\  \relax
[HC2000] 515             & 05:35:11.21 & -05:22:10.8 & M7:(SHC-ir)  \\  \relax
[HC2000] 127             & 05:35:11.33 & -05:24:26.6 & M0(SHC-ir)  \\  \relax
[HC2000] 559             & 05:35:11.37 & -05:21:54.0 & M8(SHC-ir)  \\  \relax
[HC2000] 709             & 05:35:11.63 & -05:22:46.1 & M5(SHC-ir)  \\  \relax
[H97b] 10420             & 05:35:11.67 & -05:26:08.6 & early-M(RRL)  \\  \relax
[HC2000] 708             & 05:35:11.92 & -05:22:50.9 & M4(SHC-ir)  \\  \relax
[H97b] 20182             & 05:35:12.93 & -05:24:57.6 & M5.25(RRL)  \\  \relax
[OW94] 137-532           & 05:35:13.7  & -05:25:32   & $>$M9(Luc06)$>$M9.5(WLR)   \\  \relax               [HC2000] 721             & 05:35:13.18 & -05:24:24.9 & M3.5:(SHC-ir)  \\  \relax
[H97b] 20377             & 05:35:14.76 & -05:28:31.8 & M8.5(RRL)  \\  \relax
[OW94] 152-717           & 05:35:15.2  & -05:27:17   & $>$M9(Luc06)$>$M9.5(WLR)  \\  \relax
[HC2000] 600             & 05:35:15.41 & -05:21:39.5 & M5(SHC-ir)  \\  \relax
[HC2000] 4               & 05:35:15.56 & -05:25:46.8 & M4.5(SHC)M5.5(SHC-ir)  \\  \relax
[HC2000] 565             & 05:35:16.01 & -05:21:53.1 & M8(SHC-ir)  \\  \relax
[H97b] 20295             & 05:35:16.52 & -05:26:34.4 & M8.75(RRL)  \\  \relax
[HC2000] 237             & 05:35:17.41 & -05:23:41.8 & M2e(SHC)  \\  \relax
[HC2000] 469             & 05:35:17.58 & -05:22:27.8 & M(SHC-ir)  \\  \relax
[HC2000] 162             & 05:35:17.58 & -05:24:09.0 & M5.5(SHC-ir)  \\  \relax
[HC2000] 383             & 05:35:17.84 & -05:22:58.2 & M4(SHC-ir)  \\  \relax
[HC2000] 764             & 05:35:17.97 & -05:23:53.6 & M7.5(SHC-ir)  \\  \relax
[HC2000] 594             & 05:35:18.05 & -05:21:41.2 & M7.5(SHC-ir)  \\  \relax
[HC2000] 372             & 05:35:18.08 & -05:23:01.8 & M9(SHC-ir)  \\  \relax
[OW94] 183-729           & 05:35:18.3  & -05:27:29   & $>$M9(Luc06)M8.75(RRL)$>$M9.5(WLR)  \\  \relax
[OW94] 183-419           & 05:35:18.32 & -05:24:19.9 & M2.5(SHC-ir)   \\    \relax          2MASS J05351862-0526313  & 05:35:18.63 & -05:26:31.4 & M8(RRL)M7(WLR)   \\    \relax            [OW94] 188-658           & 05:35:18.8  & -05:26:58   & $>$M9(Luc06) $>$M9.5(WLR)  \\  \relax
[HC2000] 409             & 05:35:19.04 & -05:22:50.7 & M0(SHC)  \\  \relax
[HC2000] 728             & 05:35:19.51 & -05:23:39.7 & M5.5(SHC-ir)  \\  \relax
[HC2000] 123             & 05:35:19.64 & -05:24:31.6 & M0-M5(SHC)M7.5(SHC-ir)  \\  \relax
[HC2000] 59              & 05:35:19.68 & -05:25:05.2 & K8-M3(SHC)  \\  \relax
[OW94] 196-659           & 05:35:19.6  & -05:27:00   & mid-M(RRL)$>$M9.5(WLR)  \\  \relax
[HC2000] 366             & 05:35:19.63 & -05:23:03.6 & M7.5(SHC-ir)  \\  \relax
[HC2000] 210             & 05:35:19.86 & -05:23:51.6 & $>$M6:e(SHC)M7(SHC-ir)  \\  \relax
[HC2000] 365             & 05:35:20.13 & -05:23:04.5 & M7:(SHC-ir)  \\  \relax
[HC2000] 429             & 05:35:20.64 & -05:22:41.2 & M7-M9(SHC)M7.5(SHC-ir)  \\  \relax
[HC2000] 732             & 05:35:20.77 & -05:22:39.5 & M2.5:(SHC-ir)  \\  \relax
[HC2000] 731             & 05:35:20.79 & -05:22:36.3 & K7:(SHC-ir)  \\  \relax
[H97b] 20241             & 05:35:20.90 & -05:25:34.5 & M8(SHC-ir)  \\  \relax
[HC2000] 403             & 05:35:21.02 & -05:22:54.3 & M7(SHC-ir)  \\  \relax
[HC2000] 729             & 05:35:21.12 & -05:22:50.2 & M7(SHC-ir)  \\  \relax
[HC2000] 27              & 05:35:21.29 & -05:25:33.2 & M5(SHC-ir)  \\  \relax
[H97b] 20248             & 05:35:21.35 & -05:25:35.0 & M8(SHC-ir)  \\  \relax
[HC2000] 15              & 05:35:21.61 & -05:25:40.6 & M0-M1(SHC)M3.5(SHC-ir)     \\  \relax
[H97b] 10597             & 05:35:21.69 & -05:26:52.6 & M7.75(RRL)  \\  \relax
[HC2000] 730             & 05:35:21.71 & -05:22:38.3 & M4(SHC-ir)  \\  \relax
[HC2000] 30              & 05:35:21.83 & -05:25:28.4 & M0-M3e(SHC)M2(SHC-ir)  \\  \relax
[H97b] 10605             & 05:35:21.99 & -05:24:53.3 & $<$K7(SHC-ir)  \\  \relax
[HC2000] 55              & 05:35:22.12 & -05:25:07.6 & M8(SHC-ir)  \\  \relax
[H97b] 10606             & 05:35:22.17 & -05:27:44.7 & M5.75(RRL)  \\  \relax
[HC2000] 200             & 05:35:22.54 & -05:23:54.8 & M3(SHC-ir)  \\  \relax
[HC2000] 316             & 05:35:23.34 & -05:23:20.8 & M3.5-M5e(SHC)  \\  \relax
[HC2000] 212             & 05:35:23.54 & -05:23:51.0 & M9(SHC-ir)  \\  \relax
[H97b] 10626             & 05:35:25.03 & -05:24:38.4 & M7(SHC)  \\  \relax
[HC2000] 167             & 05:35:25.27 & -05:24:06.5 & M6-M8(SHC)M7.5(SHC-ir)    \\  \relax            [HC2000] 48              & 05:35:25.54 & -05:25:11.7 & M4(SHC-ir)M6.5(WLR)  \\  \relax
[HC2000] 568             & 05:35:25.56 & -05:21:54.0 & $<$K7(SHC-ir)  \\  \relax
[HC2000] 70              & 05:35:25.67 & -05:25:02.6 & M9(SHC-ir)   \relax
\enddata

\tablenotetext{a}{ The star names are SIMBAD compatible.
Optical names are preferred to infrared names, though 2MASS is used above coordinate-based optical names.
It should be noted that the [H97b] identifiers in SIMBAD
that begin at 10000 actually originate with Herbst et al. (2002)
and those beginning with 20000 originate in Rodriguez-Ledesma et al. (2009). 
}

\tablenotetext{b}{ J2000.  Note that the reported coordinates have varying levels of precision and hence accuracy in this crowded field. }

\tablenotetext{c}{ The sources of the literature spectral types (SpT) have the same codes as in Table 2.}

\end{deluxetable}
}

\end{document}